\documentclass[
superscriptaddress,
amsmath,
amssymb,
aps,
twocolumn,
pra,
floatfix,
]{revtex4-2}

\usepackage{graphicx}
\usepackage{dcolumn}
\usepackage{bm}
\usepackage{tikz}
\usepackage{quantikz}

\usepackage[hidelinks]{hyperref}
\usepackage{orcidlink}
\usepackage{xspace}
\usepackage{siunitx}
\usepackage{cases}
\usetikzlibrary{decorations.pathreplacing}
\usepackage{float}
\usepackage[normalem]{ulem}
\usepackage{pgfplots}
\usepackage{mathrsfs}
\usepackage{standalone}
\usepackage{booktabs}
\usepackage{adjustbox}
\usepackage{tabularx}
\usepackage{hyperref}
\usepackage{afterpage}

\pgfplotsset{
    compat=newest,
    width=0.75\textwidth,
    height=0.45\textwidth,
    scale only axis,
    xlabel=$x$,
    ylabel=$y$,
    yticklabel style={text width=2em},
    grid=major,
    grid style={line width=.1pt, draw=gray!50},
}
\graphicspath{{Figures/}}

\DeclareFontFamily{OT1}{modpzc}{}
\DeclareFontShape{OT1}{modpzc}{m}{it}{<-> s*[1.15] pzcmi7t}{}
\DeclareMathAlphabet{\mathpzc}{OT1}{modpzc}{m}{it}
\newcommand{\vecc}[1]{\mbox{$\uline{#1}$}\xspace}
\renewcommand{\vec}[1]{\mbox{$\bm{#1}$}\xspace}
\newcommand{\matc}[1]{\mbox{$\uuline{#1}$}\xspace}
\newcommand{\mat}[1]{\mbox{$\bm{#1}$}\xspace}
\renewcommand{\tensor}[1]{\mbox{\pmb{{#1}}}\xspace}
\newcommand{\qtt}[1]{\mbox{{$\mathpzc{#1}$}\xspace}}
\newcommand{\physone}[1]{r_{#1}}

\newcommand{\linkone}[1]{\ensuremath{\alpha_{#1}}}
\newcommand{\linktwo}[1]{\ensuremath{\beta_{#1}}}

\newcommand{\nx}{\ensuremath{\vecc{n}^{x}_j}}
\newcommand{\nxqtt}{\ensuremath{\qtt{n}^{x}}}
\newcommand{\nz}{\ensuremath{\vecc{n}^{z}_j}}
\newcommand{\nzqtt}{\ensuremath{\qtt{n}^{z}}}
\newcommand{\dcyl}{\ensuremath{D}}

\newcommand{\nvector}[2]{$n_\eta={#1}\, \land \, n_\xi={#2}$}
\newenvironment{rcases2}{\left.\begin{aligned}}{\end{aligned}\right\rbrace}

\definecolor{red}{RGB}{228,26,28}
\definecolor{red1}{RGB}{180,26,28}
\definecolor{red2}{RGB}{255,26,28}
\definecolor{blue}{RGB}{55,126,184}
\definecolor{green}{RGB}{77,175,74}
\definecolor{purple}{RGB}{152,78,163}
\definecolor{orange}{RGB}{255,127,0}
\definecolor{yellow}{RGB}{255,255,51}
\definecolor{brown}{RGB}{166,86,40}
\definecolor{pink}{RGB}{247,129,191}
\definecolor{gray}{RGB}{170,170,170}
\definecolor{lightgray}{RGB}{230,230,230}
\definecolor{darkgray}{RGB}{120,120,120}

\begin{document}
\title{Quantum-Inspired Tensor-Network Fractional-Step Method for Incompressible Flow in Curvilinear Coordinates
}

\author{Nis-Luca~van~H\"ulst\orcidlink{0009-0004-9893-3614}}\email[Corresponding author: ]{nis-luca.van.huelst@uni-hamburg.de}
\affiliation{
Institute for Quantum Physics, University of Hamburg, Luruper Chaussee 149, D-22761 Hamburg, Germany
}
\author{Pia~Siegl\orcidlink{0000-0003-2249-8121}}
\affiliation{
Institute for Quantum Physics, University of Hamburg, Luruper Chaussee 149, D-22761 Hamburg, Germany
}
\affiliation{
Institute of Software Methods for Product Virtualization, German Aerospace Center (DLR), N\"othnitzer Straße 46b, D-01187 Dresden, Germany
}

\author{Paul~Over\orcidlink{0000-0001-7436-5254} }
\affiliation{
Institute for Fluid Dynamics and Ship Theory, Hamburg University of Technology, Hamburg D-21073, Germany
}
\author{Sergio~Bengoechea\orcidlink{0009-0001-8205-5878} }
\affiliation{
Institute for Fluid Dynamics and Ship Theory, Hamburg University of Technology, Hamburg D-21073, Germany
}
\author{Tomohiro Hashizume\orcidlink{0000-0002-7154-5417}}
\affiliation{
Institute for Quantum Physics, University of Hamburg, Luruper Chaussee 149, D-22761 Hamburg, Germany
}
\affiliation{
The Hamburg Centre for Ultrafast Imaging, Luruper Chaussee 149, D-22761 Hamburg, Germany
}

\author{Mario Guillaume Cecile \orcidlink{0000-0002-2076-6236}}
\affiliation{
Institute for Quantum Physics, University of Hamburg, Luruper Chaussee 149, D-22761 Hamburg, Germany
}
\author{Thomas~Rung\orcidlink{0000-0002-3454-1804} }
\affiliation{
Institute for Fluid Dynamics and Ship Theory, Hamburg University of Technology, Hamburg D-21073, Germany
}

\author{Dieter~Jaksch\orcidlink{0000-0002-9704-3941} }
\affiliation{
Institute for Quantum Physics, University of Hamburg, Luruper Chaussee 149, D-22761 Hamburg, Germany
}
\affiliation{
The Hamburg Centre for Ultrafast Imaging, Luruper Chaussee 149, D-22761 Hamburg, Germany
}
\affiliation{
Clarendon Laboratory, University of Oxford, Parks Road, Oxford OX1 3PU, United Kingdom
}

\date{\today}

\begin{abstract}
We introduce an algorithmic framework based on tensor networks for computing fluid flows around immersed objects in curvilinear coordinates. We show that the tensor network simulations can be carried out solely using highly compressed tensor representations of the flow fields and the differential operators and discuss the numerical implementation of the tensor operations required for computing fluid flows in detail. The applicability of our method is demonstrated by applying it to the paradigm example of steady and transient flows around stationary and rotating cylinders. We find excellent quantitative agreement in comparison to finite difference simulations for Strouhal numbers, forces and velocity fields. The properties of our approach are discussed in terms of reduced order models. We estimate the memory saving and potential runtime advantages in comparison to standard finite difference simulations. We find accurate results with errors of less than 0.3\% for flow-field compressions by a factor of up to 20 and differential operators compressed by factors of up to 1000 compared to sparse matrix representations. We provide strong numerical evidence that the runtime scaling advantages of the tensor network approach with system size will provide substantial resource savings when simulating larger systems. Finally, we note that, like other tensor network-based fluid flow simulations, our algorithmic framework is directly portable to a quantum computer leading to further scaling advantages.

\bigskip


\end{abstract}
\maketitle

\section{Introduction}
\label{sec:intro}
Engineering computational fluid dynamics (CFD) is subjected to an ever-increasing demand for higher spatial and temporal resolution. An improved dynamic resolution is crucial to understand complex multiphysics and multiscale dynamics whose analysis has so far been impossible due to a lack of computing power. The most prominent example probably relates to the simulation of turbulent flows with the full resolution of all spatiotemporal scales in a direct numerical simulation (DNS).
This requirement can pose enormous computational challenges because the range of relevant scales increases nonlinearly with the Reynolds number, which renders DNS infeasible for engineering CFD~\cite{MOSER2023,Nasa2030,Moin2007}.

Since the ultimate goal of resolving turbulence by DNS remains elusive in CFD, approximate methods based on modeling turbulence physics, coarse-grained or reduced order models (ROMs) have been developed. Turbulence closure models ~\cite{WILCOX06,LAUNDER83,Pope2000} based on empirical parameters
substantially reduce the resolution requirements, but at the expense of a lower generality and less detailed results.
Alternatively, ROMs~\cite{Aubry1991, Sirovich1987, Taira2017, Schmid2010, Kutz2016, Brunton2020, Brunton2022, Pfeffer2023, Ramezanian2021}, utilize scale-resolving strategies that mitigate the curse of dimensionality problem~\cite{Bellman2010} by reducing state space complexities~\cite{Berkooz1993, Holmes1996, Holmes1997, Kunisch1999, Rowley2005, Venturi2006}, 
for example, through modal decomposition.
 However, CPU-efficient ROMs do not provide reliable (long-term) predictions due to the simplification of the dynamics.

Advances in CFD may also arise from the development of novel computing hardware~\cite{Nasa2030}. In this regard, quantum computers (QCs) have the potential to exponentially reduce the amount of memory required for storing highly resolved discretized flow fields~\cite{Givi2020, Jaksch2023a}.
Recent overviews of the prospects of different quantum-based strategies in engineering applications, especially CFD, are provided in~\cite{Bharadwaj2020,Jaksch2023a,Riofrio2024}.
Aiming to apply quantum mechanical concepts, it is important to understand its major bottleneck. This arises from the linear nature of quantum mechanics, and poses a challenge for simulating nonlinear differential equation problems on a QC. Other challenges relate to the measurement of quantum (flow) states as well as hardware-related influences of noise and decoherence.

Proposals for combining the memory advantage with quantum algorithms to solve nonlinear CFD problems 
are, e.g.,~based on Carleman linearization~\cite{Sanavio24,Sanavio25}, efficient quantum algorithms for solving linear(ised) equation systems~\cite{Harrow2009, Childs2017} or 
quantum realizations of mesoscopic lattice Boltzmann frameworks~\cite{Budinski2021, Budinski2022, Itani24, Wawrzyniak2024}. 
Other options refer to transforming the nonlinear differential flow equations into equivalent Schr\"odinger equations, cf.~\cite{Madlung27}, whose dynamics can be conveniently simulated on a QC using a Hamiltonian simulation algorithm~\cite{Meng23,Meng24,Salasnich24,Brearley2024,Over2024b}

To balance unfavorable properties of current quantum hardware, the use of hybrid quantum-classical computers can be advantageous.
Related approaches often transform the governing transport equations into an optimization problem and parameterize the state. The parameters encode the nonlinearities and are optimized by a corresponding optimization procedure.
Such hybrid quantum-classical variational quantum methods~\cite{Peruzzo_2014,Cerezo2021,Bharadwaj23} 
benefit from the exponential memory saving and might also achieve a runtime advantage, see, e.g.,~\cite{Lubasch2020, Jaksch2023a, Over2024a, Bengoechea2024} for CFD applications.
\bigskip

Another avenue is to reformulate classical algorithms using quantum computing strategies to develop quantum-inspired methods that potentially offer higher computational power.
Such quantum-inspired methods can be implemented on classical computers but may also use quantum computers as accelerators.
Their algorithms often utilize tensor networks~\cite{Schollwoeck2011} as a programming paradigm for the quantum part of the computation~\cite{Termanova2024, Siegl2025}.
Since sufficiently advanced quantum hardware is not yet available, their performance and accuracy is often analyzed and benchmarked on classical hardware.
Fully scale-resolving  quantum-inspired tensor network  simulations are 
computationally attractive only
if the entanglement -- i.e.,~the internal correlations between different parts of the system -- is limited, effectively turning them into ROMs~\cite{Jaksch2023a,Garcia2024}. As conventional intrusive ROMs~\cite{Rozza2022}, these methods allow complete CFD simulation within a compressed subspace, akin to studies in quantum many-body systems~\cite{Eisert2010}. 

Quantics tensor trains~\cite{Khoromskij2011}, also known as a special type of matrix product states (MPS) in quantum physics, and hereafter referred to simply as tensor trains (TTs), have emerged as strong candidates for quantum-inspired tensor network CFD methods. Building on the pioneering work of~\citeauthor{Dolgov2012}~\cite{Dolgov2012}, which utilized the TT format to solve semi-discrete parabolic problems, recent studies have successfully extended quantum-inspired ROMs to a broad variety of increasingly advanced CFD problems~\cite{Dolgov2012,Richter2021,Lubasch2018,Hoelscher2024,Kiffner2023,Ye2024,Peddinti2024,Michailidis2024,Danis2025,Pisoni2025,Lively2025,Arenstein2025}. They mainly utilize conventional finite difference (FD)  frameworks and address various challenges faced by conventional CFD, such as incorporating higher-order spatial and temporal approximation schemes.
Through this approach, significant speedups have been reported: a 12-fold reduction in computational time for simulating a 2D decaying-turbulence flow on graphic-cards~\cite{Hoelscher2024}, one order of magnitude runtime improvement~\cite{Kiffner2023} compared to classical FD for the paradigmatic 2D lid-driven cavity problem and~\citeauthor{Gourianov2024}~\cite{Gourianov2024} reported a reduction of memory and computational cost up to six and three orders of magnitude, respectively, for the (5+1)D joint probability density-function simulation of a chemically reactive turbulent flow.

Applications of TT methods to complex geometries are still in their infancy and simulations around non-rectangular bodies represent a challenge. Nowadays, mesh-based classical CFD methods predominantly refer to  body-fitted, unstructured polyhedral meshes~\cite{DEMIRDZIC1995,OFOAM1998}.
Other approaches employ the immersed boundary method (IBM) to capture complex geometries usually  embedded in structured Cartesian grids~\cite{Lai2000,Peskin2002,TAIRA2007,Verzicco2023}.
In addition, particularly in the area of complex multi-body flow  simulations, there are approaches based on overlapping domain-specific grids, which are structured and body-fitted. A key challenge of overset grids is certainly the organization of the dynamic connectivity between grids and the related conservative inter-grid interpolation~\cite{VOLKNER2017,SHARMA2021}.
To date, only two studies have investigated TT-based CFD methods for complex domains. The first one was published by~\citeauthor{Peddinti2024}~\cite{Peddinti2024}. It employs a direct-forcing IBM approach on a Cartesian mesh to simulate flow around circular and square cylinders, as well as a NACA0042 airfoil. However, the added geometry-representing penalty terms inherent to this method generally introduce extra algorithmic complexity and computational overhead.
The second one, reported by~\citeauthor{Kornev2023}~\cite{Kornev2023}, utilizes block-structured Cartesian meshes to simulate flow in a T-shaped channel, yet can not resolve non-rectangular objects.
\bigskip

The present paper reports the implementation of a TT-based CFD procedure for body-fitted structured-grid discretizations using curvilinear coordinates. 
The transformation of Cartesian differential operators into curvilinear operators within the TT format will be outlined in detail.
Analogous to classical approaches, the procedure is based on a fractional-step algorithm~\cite{Chorin1968,Kim1985,Patankar2018}.
This enables an accurate and efficient simulation of incompressible flows around submerged objects, extending the applicability of TT methods beyond Cartesian grids. 
An investigation of overset grids is,  however, beyond the scope of this article but it paves the way for future TT applications on these type of grids.

This publication is structured as follows: Section~\ref{sec:QTT_Principles} introduces the TT format and discusses the realization and scaling of basic algebraic operations such as addition and multiplication of fields and operators, together with a short introduction into the density matrix renormalization group algorithm used to solve linear systems of equations. The subsequent Section~\ref{sec:comp_model} sketches a generic physical model of an incompressible flow around an obstacle and introduces a discrete curvilinear coordinate transformation. Furthermore, it covers the representation of the flow fields, grid generation, and the construction of curvilinear differential operators, all within the TT format, as well as the implicit–explicit scheme for time integration.

In Sections \ref{subsec:Steady}–\ref{sec:rot_cylinder} the proposed framework is applied to two-dimensional, laminar flows around both non-rotating and rotating circular cylinders in steady and transient regimes ($20\leq Re \leq 200$).
We observe excellent prediction and rapid convergence of the TT algorithm toward the reference values characterizing the system dynamics, such as mean lift coefficients and Strouhal numbers. 
Remarkably, for the steady (non-rotating) cylinder case the proposed TT solver achieves a relative error below $0.3\%$ in the velocity magnitude utilizing only 5.8\% degrees of freedom of classical simulations, i.e.,~a 20-fold compression. The flow dynamics in the transient case are also accurately predicted with an absolute error below $1 \times 10^{-3}$ in the Strouhal number at compression ratios as low as 18\%. Section~\ref{sec:Performance} discusses the computational performance of the TT solver on the transient flow case. We demonstrate that the computational effort required to refine the discretization is significantly reduced when using the TT method compared to classical FD simulations, particularly in cases where the bond dimension does not grow substantially with the system size. 
Furthermore, we show that the required curvilinear differential operators admit a compression of up to 1000-fold in the TT format, efficiently encoding the orthogonal cylindrical grid.
The last section, Section~\ref{sec:conclusions}, is devoted to conclusions and outlook, where we discuss possible implications for future developments of tensor-network-based fluid dynamics.

To accommodate a broad audience, this publication makes extensive use of Appendices to provide additional technical details and background. 
Appendix~\ref{App:DiscretizationCurvis} provides additional information regarding the discretization and the curvilinear coordinate transformation. Appendix~\ref{App:MatrixVectorSorting} introduces the vectorization of a matrix and Appendix~\ref{App:Readout} specifies how rows and columns are extracted from the TT representation of a vectorized matrix. The decomposition of a vector into a TT is explained in detail in Appendix~\ref{App:TensorTrain}. The grid points used in the simulations and the details on the force computation are specified in Appendix \ref{app:gridpoints} and \ref{app:stresses}, respectively. Appendix~\ref{app:AlgorithmModificationsforTransient} introduces the algorithmic modifications necessary to compute the transient flow and Appendix~\ref{app:openfoam} discusses the verification of the in-house FD solver using an OpenFOAM simulation for comparison.

The publication employs lower-case Latin letters to describe field properties and uppercase Latin letters 
indicate reference quantities used to non-dimensionalize 
the field quantities. Dimensional field quantities are denoted by a tilde. In a symbolic representation, the number of underlines indicates the tensorial order, e.g.,~$\underline{\tilde x}$ for the position vector and $\underline \nabla$ for the non-dimensional spatial derivative vector. Spatial vectors and tensors are usually defined with reference to Cartesian coordinates.
Furthermore, the notation for discretized quantities follows
\begin{equation*}
\begin{split}
     \text{Vectors (order-1 tensor):}& \,\vec{v}\,,\\
    \text{Matrices (order-2 tensor):}&\, \mat{A}\,,\\
    \text{Order-3 or higher tensors:}&\, \tensor{v},\, \tensor{A}\,, \\ 
    \text{Tensor Trains:}&\,  \qtt{v}, \,\qtt{A} \, .
\end{split}
\end{equation*}

\section{Tensor Trains: Principles and Fundamental Operations}
\label{sec:QTT_Principles}
This section presents a reduced order modeling framework based on TTs in their binary (dyadic) form -- usually called quantics tensor trains or understood as a special case of matrix product states -- which we adopt for its compatibility with quantum hardware. Throughout the remainder of the paper, the term TT will always refer to this binary specialization.

We start by defining the notation and describe the encoding of high-dimensional tensors, representing state vectors and operators, into the TT format. Subsequently, the algebraic operations required in this framework are detailed in Section~\ref{subsec:QTT:algebraicoperations}. 
Moreover, the complexity of the operations is discussed in Section~\ref{Sec:algorithmic_scaling}. 
Throughout this section, we practice the Einstein summation convention, where repeated indices imply summation.
For reordering vectors to matrices and vice versa, the index fusion/extension always assumes a column ordering. In general, it must be chosen consistently, and it mostly depends on the programming language, which is in our case \textsc{julia} \cite{bezanson2017}. For the sake of understandability, the sorting approach is detailed in the Appendix~\ref{App:MatrixVectorSorting}. For any contraction, we rely on a toolbox determining the most efficient order.

\smallskip

\begin{figure}[htbp]
    \centering
    \includegraphics[width=0.9\linewidth]{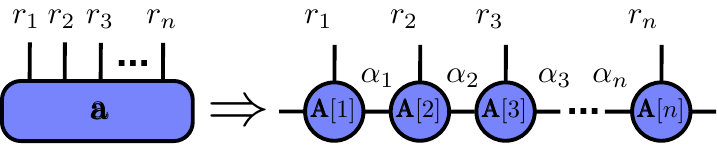} 
    \caption{Example for the decomposition of an arbitrary vector $\vec{a}$, reshaped as $2^n$-component tensor $\tensor{a}_{\physone{1}, \physone{2}, \ldots, \physone{n}}$, into the TT format by performing a series of SVDs, where $\tensor{A}[k]$ are order-3 tensors with mode indices $\physone{k}\in \{0,1 \}$, and bond indices $\linkone{k}$. The corresponding mathematical interpretation is given in Eq.~\eqref{eq:simpleMPS}.}
    \label{fig:QTT_vec}
\end{figure}

For an introduction to the fundamentals of TTs, a {$2^n$}-component vector is reshaped into an order-$n$ tensor $\tensor{a}_{\physone{1}, \physone{2}, \ldots, \physone{n}}$ with binary $\physone{k} \in \{0, 1 \}$ for $k=1,\ldots, n$ and subsequently decomposed into a TT representation of $n$ order-3 tensors $\tensor{A}[k]$ (TT-cores)~\cite{Kornev2023,Oseledets2011,Oseledets2009, Khoromskij2014}, as it is schematically depicted in Fig.~\ref{fig:QTT_vec}. For the representation of tensors in this work, we follow the (graphical) \textit{Penrose} notation~\cite{Penrose1971}, where a vector is an object with one leg (cf.~Fig.~\ref{fig:QTT_vec}), a line sticking out of the blue box/circle, a matrix has two legs, and an order-$n$ tensor shows $n$ legs. The number of outgoing legs thereby determines the order of the tensor. A link between any two tensors, indicated with a lowercase Greek letter, is referred to as a bond (or artificial) index implying summation. The outer indices instead are called physical (or real) indices with dimension $2$
and indicated by a lowercase {Latin} letter. 
\newline
The decomposition illustrated in Fig.~\ref{fig:QTT_vec} is performed by the successive application of singular value decompositions (SVDs). As an alternative, the TT-cross algorithm~\cite{OSELEDETS2010TTCROSS, Ritter24} can be employed, or for certain cases, the TT-vectors can be derived analytically, as is the case of {polynomial and} trigonometric functions~\cite{Oseledets2012}. For further insights, the interested reader is referred to the Refs.~\cite {Schollwoeck2011,Orus2013}. Additionally, an exemplary analytical tensor decomposition is given in Appendix~\ref{App:TensorTrain}. In any case, this decomposition is not unique, and a coherent order within a so-called canonical form~\cite{Orus2013, Schollwoeck2011} is recommended. Within the toolbox we maintain a right-canonical form. 

The dimension of the bond indices $\alpha_k$ connecting individual order-3 tensors can be set by either neglecting singular values below a certain threshold or by manually selecting the first $\chi$ diagonal entries in the singular value matrix~\cite{Schollwoeck2011}. In this regard, the complete TT-vector reads as
\begin{equation}
 \qtt{a} : = 
\tensor{A}[1]_{\linkone{0}, \linkone{1}}^{\physone{1}}\tensor{A}[2]_{ \linkone{1}, \linkone{2}}^{\physone{2}} \cdots \tensor{A}[n]_{\linkone{n-1}, \linkone{n}}^{\physone{n}}\, {\simeq} \ \tensor{a}_{\physone{1}, \physone{2}, \ldots, \physone{n}} \, ,
\label{eq:simpleMPS}
\end{equation}
where the dimensions of bond indices $\linkone{k}$ follow $\text{dim}(\linkone{k}) = \text{min}(2^k,2^{n-k}, \chi)$ for $k=0,\ldots, n$. Accordingly, $\chi$ is always referred to as the maximal value in the set $\{\text{dim}(\alpha_0),\text{dim}(\alpha_1),\hdots, \text{dim}(\alpha_n)\}$, and controls the expressiveness of the representation, possibly affecting its accuracy. In the case where the bond dimension is not under truncation, the TT representation $\qtt{a}=\tensor{a}_{\physone{1}, \physone{2}, \ldots, \physone{n}}$ is exact.

\begin{figure}[htbp]
    \centering
    \includegraphics[width=0.9\linewidth]{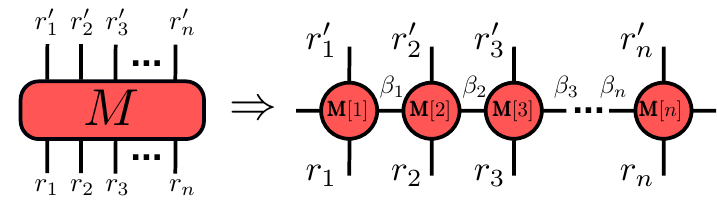}
    \caption{Graphical depiction of the TT representation $\qtt{M}$ of a  matrix $\mat{M}$.  The red circles with outgoing legs denote TT-cores, where $(\physone{1}',...,\physone{n}')$ represents the rows of $\mat{M}$ and $(\physone{1},...,\physone{n})$ enumerates its columns. The corresponding mathematical description is given in Eq.~\eqref{eq:simpleMatrix}.}
   \label{fig:QTT_mat}
\end{figure}

Similarly, a matrix $\mat{M}$ can be represented as a TT-matrix~$\qtt{M}$ (TT-operator) by local order-$4$ tensors $\tensor{M}[k]$ with incoming (not primed) and outgoing (primed) modes of the same mode size (square matrices), $2$, such that the contraction of incoming mode indices with the mode indices of a TT-vector naturally gives a new TT-vector. The TT-matrix is illustrated in Fig.~\ref{fig:QTT_mat}.

The matrix $\mat{M}$ can be encoded through the successive application of SVDs, defining the maximal bond dimension $\chi(\qtt{M})$ as the largest value in the set $\{\text{dim}(\beta_0), \text{dim}(\beta_1), \dots, \text{dim}(\beta_n)\}$. The TT cross algorithm~\cite{OSELEDETS2010TTCROSS} may also be employed to further reduce computational effort when obtaining the TT-matrix. However, in most cases, numerical decompositions are not required, as the TTs in these instances can be derived analytically~\cite{Kazeev2012}.

Each Latin index again holds a mode size of $2$. The resulting TT-matrix~\cite{Kornev2023}, consisting of order-$4$ tensors, is given by
\begin{equation}
    \qtt{M} : = 
    \tensor{M}[1]_{\linktwo{0}, \linktwo{1} }^{\physone{1}', \physone{1}}\tensor{M}[2]_{ \linktwo{1} , \linktwo{2} }^{\physone{2}', \physone{2}} \ldots \tensor{M}[n]_{\linktwo{n-1} , \linktwo{n}}^{\physone{n}', \physone{n}}  \simeq \tensor{M}_{\physone{1}, \physone{2}, \ldots, \physone{n}}^{\physone{1}', \physone{2}', \ldots, \physone{n}'}\,.
    \label{eq:simpleMatrix}
\end{equation}
Again, without truncation, the above representation is exact.

\subsection{Algebraic Operations}
\label{subsec:QTT:algebraicoperations}
In this TT-representation of matrices and vectors, the common algebraic operations needed for solving PDEs, can be performed efficiently. These include vector-vector addition (e.g.,~$\qtt{a}+\qtt{b}$), matrix-vector product (e.g.,~$\qtt{M} \qtt{a})$, and elementwise multiplication of two vectors (e.g.,~$\qtt{a} \odot \qtt{b})$, and are introduced in the following Sections~\ref{subsubsec:vect-vect-add}-\ref{sec:QTT:Elementwise}. Solving linear systems of equations is done variationally, as explained in Section~\ref{sec:QTT:SLE}.
Application of TT operations may result in an increased bond dimension, which is typically controlled through regular truncation of the TT-vector (TT-rounding)~\cite{Oseledets2011,Schollwoeck2011} to maintain computational efficiency.

\subsubsection{Vector-Vector Addition}
\label{subsubsec:vect-vect-add} 
Given two TT-vectors $\qtt{a}$ and $\qtt{b}$, with cores $\tensor{A}[k]^{\physone{k}}$ and $\tensor{B}[k]^{\physone{k}}$ and bond dimension $\chi(\qtt{a})$ and $\chi(\qtt{b})$, a summation of $\qtt{a}$ and $\qtt{b}$ results in the new TT-vector $\qtt{c}$. For the summation, each core $C[k]$ of $\qtt{c}$ is built as 
\begin{equation}
\tensor{C}[k]^{\physone{k}}=\begin{bmatrix}
\tensor{A}[k]^{\physone{k}} & 
{\pmb{0}} \\
{\pmb{0}} & \tensor{B}[k]^{\physone{k}}
\end{bmatrix} \, = \tensor{A}[k]^{\physone{k}} \oplus \tensor{B}[k]^{\physone{k}} \, , 
\end{equation}
where the direct sum $\oplus$ is the shorthand notation for this block encoding. For $k=1$ ($k=n$), the TT-cores of $\qtt{a}$ and $\qtt{b}$ are concatenated in a row (column) vector, respectively. The resulting TT-vector $\qtt{c}$ will have a bond dimension $\chi(\qtt{c})=\chi(\qtt{a})+\chi(\qtt{b})$, which may be truncated.

\subsubsection{Matrix-Vector Product}
\label{subsubsec:mat-vect-prod}
\begin{figure}[htbp]
    \centering
    \includegraphics[width=0.9\linewidth]{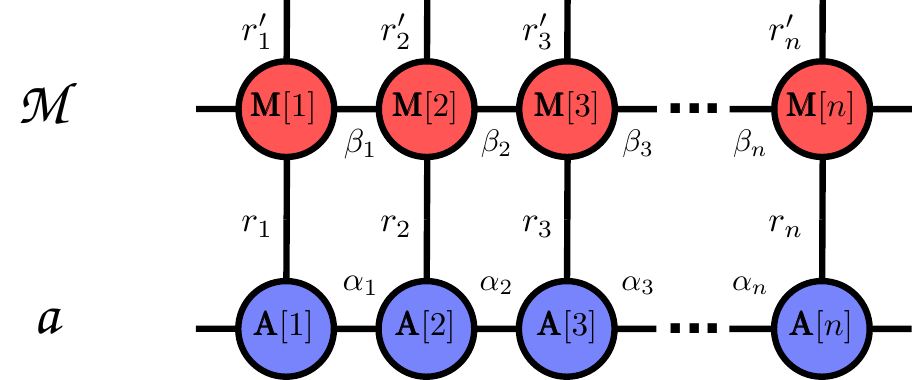}
    \caption{Graphical representation of TT-matrix acting on a TT-vector, which is achieved by contraction over common physical indices.}
    \label{fig:QTT_mat_vec_op}
\end{figure}

The matrix-vector product is realized as a combination of TTs~\cite{Kazeev2012}. For the application of a TT-matrix $\qtt{M}$ to a TT-vector $\qtt{a}$, multiple contractions along the physical indices are required, as it is depicted in Fig.~\ref{fig:QTT_mat_vec_op}. 
In mathematical terms, the TT-vector $\qtt{b} = \qtt{M} \qtt{a}$ has the cores
\begin{equation}
    \tensor{B}[k]_{1, (\linkone{k} \linktwo{k})}^{\physone{k}'} = \sum_{r_k} \tensor{M}[k]_{1, \linktwo{k}}^{\physone{k}', \physone{k}} \tensor{A}[1]_{1, \linkone{k}}^{\physone{k}}
\end{equation}
such that the resulting tensor train is
\begin{equation}
    \qtt{b}= \tensor{B}[1]_{1, (\linkone{1} \linktwo{1})}^{\physone{1}'}\tensor{B}[2]_{ (\linkone{1} \linktwo{1}), (\linkone{2} \linktwo{2})}^{\physone{2}'} \ldots \tensor{B}[n]_{ (\linkone{n}\linktwo{n}), 1}^{\physone{n}'} \, .
\end{equation}

Here, $(\linkone{k}\, \linktwo{k})$ denotes the fused bond index formed by combining $\linkone{k}$ and $\linktwo{k}$, with total dimension $\text{dim}(\linkone{k}) \times \text{dim}(\linktwo{k})$ (Appendix~\ref{App:MatrixVectorSorting}).
As a result, this increases the bond dimension of the resulting TT-vector, which requires regular truncation (e.g.,~using SVD) for operations to remain efficient.

\subsubsection{Vector-Vector Elementwise Multiplication}
\label{sec:QTT:Elementwise}
\begin{figure}[htbp]
    \centering
    \includegraphics[width=0.95\linewidth]{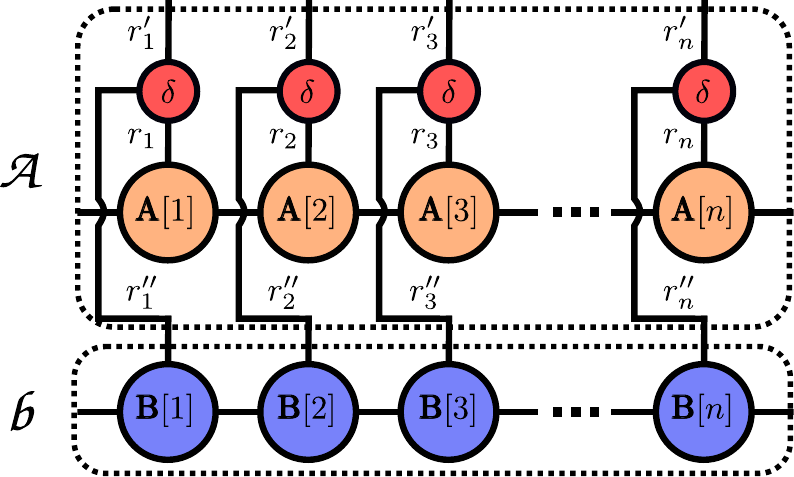}
    \caption{Graphical representation of vector-vector elementwise multiplication between two TT-vectors $\qtt{a}$ and $\qtt{b}$.}
    \label{fig:vect-vect-elementwise-mult}
\end{figure}

As the last fundamental operation, the vector-vector elementwise multiplication, indicated by $\odot$, is introduced. 
For three TT-vectors $\qtt{a}$, $\qtt{b}$, and $\qtt{c}$, the elementwise product $\qtt{c} = \qtt{a} \odot \qtt{b}$ is computed in two steps. 
The first step involves contracting the copy tensor $\pmb{\delta}^{\physone{k}'}_{\physone{k},\physone{k}''}$ tensor, defined as~\cite{Biamonte_2011}
\begin{equation}
\pmb{\delta}^{\physone{k}'}_{\physone{k}, \physone{k}''} = \begin{cases} 
1 & \text{if } \physone{k} = \physone{k}'= \physone{k}'' \\
0 & \text{otherwise}
\end{cases}
\end{equation}
into each TT-core of $\qtt{a}$, which effectively doubles its physical dimension (from $2$ to $4$). This turns the TT-vector $\qtt{a}$ into a diagonal TT-matrix $\qtt{A}$ with the vector entries along its diagonal (cf.~Appendix~\ref{App:MatrixVectorSorting}). Secondly, the resulting \text{TT} $\qtt{c}=\qtt{A}\qtt{b}$ is computed by the matrix-vector product, as it is described in Section~\ref{subsubsec:mat-vect-prod}. Both steps of the vector-vector elementwise operation are illustrated in Fig.~\ref{fig:vect-vect-elementwise-mult}. This increases the bond dimension of the resulting TT-vector according to the matrix–vector product detailed in the previous subsection.

\subsubsection{Solving Linear Systems of Equations}
\label{sec:QTT:SLE}
The discretization of partial differential equations typically leads to linear systems of equations (LSEs). In this section, we introduce a TT-based approach for solving such systems.
In classical computing, the unknowns of an LSE can be determined in many ways (e.g.,~Gauss-elimination, Cholesky decomposition, $\dots$). Within the TT format, however, one typically employs dedicated solvers; examples are described 
in Refs.~\cite{Dolgov2013,Holtz2012,Oseledets2012b,Dolgov2014}.
In the present work we adopt a single-site version of the variational density matrix renormalization group (DMRG) algorithm, also known as the alternating least-squares method, as detailed, for example, in Refs.~\cite{Holtz2012,Oseledets2012b}.

\bigskip

Generally, given TT representations of a matrix $\qtt{M}$, a vector $\qtt{a}$ and a right-hand side (RHS) $\qtt{b}$ the linear problem reads as
\begin{equation}
\label{eq:LSE}
     \qtt{M}\qtt{a}= \qtt{b} \, .
\end{equation}
For application of the DMRG, we define a cost function $\mathcal{C}(\qtt{a})$, whose global minimum corresponds to the exact solution of Eq.~\eqref{eq:LSE} if the matrix $\qtt{M}$ is symmetric and positive-semidefinite~\cite{Dolgov2012,Grossmann2007}. If this condition is not satisfied, this cost function has still been shown to perform well~\cite{Dolgov2012}; alternatively, a least-squares formulation may be employed~\cite{Grossmann2007}. To approximate the solution of the given problem, we solve the optimization problem
\begin{equation}
\label{eq:}
\min \limits_{\qtt{a}} \mathcal{C}(\qtt{a}):=  \frac{1}{2} \left(\qtt{a} \qtt{M}\qtt{a} \right) -  \qtt{a} \qtt{b} \,, 
\end{equation}
where the unknown TT-vector $\qtt{a}$ is fully parameterized. Taking the derivative of the cost functional w.r.t.~the individual TT-cores of $\qtt{a}$ reads
\begin{equation}
    \begin{split}
        \frac{\partial \mathcal{C}}{\partial \tensor{A}[k]} &=  \frac{1}{2}\frac{\partial \left(\qtt{a} \qtt{M} \qtt{a}\right)}{\partial \tensor{A}[k]} - \frac{\partial (\qtt{a}\qtt{b})  }{\partial \tensor{A}[k]} \\
            &= \mat{M}_{(\linkone{k-1}'\physone{k}' \linkone{k}'), (\linkone{k-1}\physone{k} \linkone{k})}^{\text{eff}} \tensor{A}[k]_{(\linkone{k-1}\physone{k} \linkone{k})} \\ &\quad - \vec{b}_{(\linkone{k-1}'\physone{k}'\linkone{k}')}^{\text{eff}}\, , 
    \end{split}
    \label{eq:cost_}
\end{equation}
where $(\linkone{k-1} \physone{k}\linkone{k})$ denotes the fused bond index formed by combining indices $\linkone{k-1}$, $\physone{k}$, and $\linkone{k}$ (cf.~Appendix~\ref{App:MatrixVectorSorting}).
\begin{figure*}
    \centering
        \includegraphics[width=0.75\textwidth]{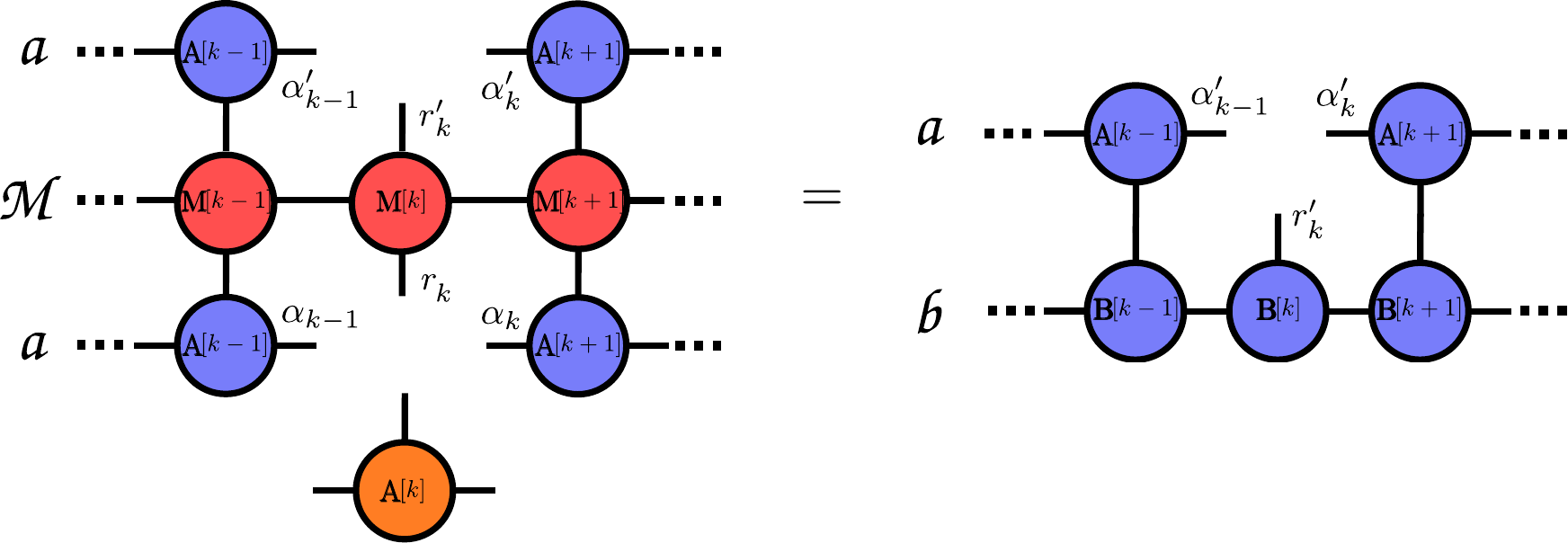}
    \caption{Graphical representation of the local linear system for the $k$-th TT-core $\tensor{A}[k]$, obtained by enforcing the stationarity condition $\partial\mathcal{C}/\partial\tensor{A}[k]=0$. Solving this system yields the updated core $\tensor{A}[k]$ (orange node).
}
    \label{fig:DMRG}
\end{figure*}
The algorithm is then realized via successive DMRG-style sweeps over all TT-cores \cite{Schollwoeck2011,Holtz2012}.  
During the update of core \(\tensor{A}[k]\), all other cores are held fixed and a small linear system for the optimal $\tensor{A}[k]$ is obtained by enforcing the stationarity condition $\partial\mathcal{C}/\partial\tensor{A}[k]=0$. This step is graphically depicted in Fig.~\ref{fig:DMRG}. The local effective matrix $\mat{M}^{\text{eff}}$ is of size $2\chi^2 \times 2\chi^2$, so that the matrix-vector product $\mat{M}^{\text{eff}} \tensor{A}[k]$ scales as $\mathcal{O}(\chi^4)$.

An alternative approach is to emulate the local matrix-vector product by keeping the constituent tensors of $\mat{M}^{\text{eff}}$ and applying successive tensor contractions~\cite{Dolgov2012}. This is reported to scale as $\mathcal{O}\big( \chi(\qtt{M})\chi^3+ \chi(\qtt{M})^2 \chi^2\big)$. However, numerical studies have shown poor convergence in the latter scheme in the case of ill-conditioned systems, rendering preconditioners mandatory. A strategy to find preconditioners directly in the TT format is outlined in~\cite{Dolgov2012}.

\subsection{TT Complexity}
\label{Sec:algorithmic_scaling}
The operations introduced in the former Sections~\ref{subsubsec:vect-vect-add}-\ref{sec:QTT:SLE}, come at certain computational expenses. 
The alternative implementation to the TT operations discussed above is the use of variational methods~\cite{Paeckel2019}, which avoid the need for intermediate truncation by directly optimizing the solution within the desired bond dimension.
A quantitative comparison between both is given in Table~\ref{tab:complexity}~\cite{Kornev2023,Kiffner2023}, where it is found that variational approaches provide better scaling with increasing bond dimension $\chi$.
Throughout our simulations, we employ the variational technique for the most computationally demanding operations, such as elementwise multiplications and the solution of LSEs.
For operations with comparatively negligible computational cost, direct methods are preferred, as they avoid the overhead associated with variational approaches and their potential convergence issues in the presence of strong numerical fluctuations. Within the literature, one finds the TT multiplication algorithm~\cite{Michailidis2024}, or the zip-up~\cite{Stoudenmire_2010}, which can improve the initially mentioned scalings.
\begin{table}[!htbp]
    \centering
        \caption{Computational complexity of the TT operations introduced in Section~\ref{sec:QTT_Principles}, including subsequent truncation to a target bond dimension (TT-rounding), and variational complexities as discussed in Ref.~\cite{Paeckel2019}.}
    \resizebox{\columnwidth}{!}{
    \begin{tabular}{@{}lll@{}}
        \toprule  
        Operation & Direct approach & Variational   \\
        \midrule  
        $\qtt{c}=\qtt{a} + \qtt{b}$ & $\mathcal{O}\big(
    (\chi(\qtt{a})+\chi(\qtt{b}))^3\big)$ &  
        $\mathcal{O}\big(\chi(\qtt{c})(\chi(\qtt{a})+\chi(\qtt{b}))^2\big)$ \\
        $\qtt{c}=\qtt{a} \odot \qtt{b}$ & $\mathcal{O}\big(\chi(\qtt{a})^3\chi(\qtt{b})^3 \big)$ &  
        $\mathcal{O}\big(\chi(\qtt{a}) \chi(\qtt{b})^2\chi(\qtt{c})$ \\
                           & 
                           & \hspace{1ex}
                        $+\chi(\qtt{a})^2 \chi(\qtt{b})\chi(\qtt{c})\big)$ \\
        \\
        mat-vec $\qtt{b}=\qtt{M} \qtt{a}$ & $\mathcal{O}\big(\chi(\qtt{M})^3 \chi(\qtt{a})^3\big)$ &  
        $\mathcal{O}\big(\chi(\qtt{M}) \chi(\qtt{a})^2\chi(\qtt{b})\big)$ 
        \\
        LSE $\qtt{M} \qtt{a}=\qtt{b}$ &  N/A
                        &  $\mathcal{O}\big( \chi(\qtt{M})\chi(\qtt{a})^3$ \\
                        &  &\hspace{1ex} $+ \chi(\qtt{M})^2 \chi(\qtt{a})^2\big)$   
                                                      \\
        \bottomrule
    \end{tabular}
    }
    \label{tab:complexity}
\end{table}

\section{Computational Model} \label{sec:comp_model}
This section briefly outlines the governing equations (Section~\ref{sec:gov}) and the classical FD-based discretization strategy based on curvilinear coordinates (Section~\ref{sec:disc}). In the latter, we describe how this discretization is implemented within the TT format, i.e.,~encoding two-dimensional scalar fields (\ref{subsub:QTT:fluid_field_enc}), generating a body-fitted grid (\ref{sec:QTT:mesh_gen}), and transforming the corresponding differential operators (\ref{subsub:QTT:Curvis}). The section concludes with detailing the fractional step method employed to solve the incompressible Navier–Stokes equations (INSE) in Section~\ref{sec:algorithm}.
\bigskip 

\subsection{Governing Equations}
\label{sec:gov}
The two‐dimensional, laminar fluid-dynamics problem is governed by the INSE for
a Newtonian fluid with constant density $\tilde{\rho}$.
Using non-dimensional spatial coordinates $\vecc{x} = \tilde{\vecc{x}}/L$, velocities   $\vecc{v} = \tilde{\vecc{v}}/U_\infty$ in addition to a non-dimensional time $t= \tilde t/T_\text{ref}$ and  pressure $p = \tilde p/P_\text{ref}$, the system of momentum (\ref{eq:Navier-Stokes-Eq-B}) and continuity (\ref{eq:Conti-Eq-B}) equations inside a physical domain $\Omega$ read
\begin{align}
    \label{eq:Navier-Stokes-Eq-B}
    \begin{split}
         \frac{\partial \vecc{v}}{\partial t}+ \big(\vecc{v} \cdot \vecc{\nabla} \big)\vecc{v} = - \vecc{\nabla} \,p + \frac{1}{Re} 
         \Delta \,\vecc{v} & + \vecc{f} \hphantom{0} 
    \end{split} \, \text{in  }\,\Omega\, , \\
    \label{eq:Conti-Eq-B}
     \vecc{\nabla} \cdot \vecc{v}   &= 0 \hphantom{\vecc{f}}  \, \text{in }\, \Omega\, ,
\end{align}
where $\vecc{\nabla}$ denotes the vector of the non-dimensional spatial derivatives, 
$Re=U_\infty L/\tilde\nu$
is the Reynolds number,   $T_\text{ref} = L/U_\infty$ is a convective reference time, $U_\infty$ is the reference (approach flow) velocity, $\tilde \nu$ refers to the (constant) kinematic viscosity of the fluid and
$L$ denotes a reference length, typically the diameter $D$ of the investigated cylinder studied in this paper, cf.~Fig.~\ref{fig:cylinder}. The employed reference pressure refers to  $P_\text{ref}=\tilde \rho U_\infty^2$
and $\vecc{f} = \vecc{\tilde f} L/ U_\infty^2$ denotes a non-dimensional volume-specific body force.

The boundary of the object, denoted $\Gamma$, is subject to Dirichlet (no-slip) conditions $\vecc{v}_\Gamma$ for the velocity and a natural (Neumann) condition for the pressure. The exterior boundaries of the domain $\partial\Omega$ are assigned to Dirichlet conditions  for both velocity and pressure, viz. 
\begin{equation}
\begin{alignedat}{4}
  \vecc v - \vecc v_\Gamma \; &= \; \vecc 0\ , 
    &\quad \quad \vecc\nabla p\cdot\vecc n \; &= \; 0\ ,
    &\quad &\text{on }\Gamma \ , \\[4pt]
  \vecc v - \vecc v_\infty  \; &= \; \vecc 0\ , 
    &\quad p                 \; &= \; 0\ ,
    &\quad &\text{on }\partial\Omega \, .
\end{alignedat}
\label{eq:bound}
\end{equation}
Initial conditions are prescribed by a spatially dependent $\vecc{v} = \vecc{v}_\text{init.}$ for $t=0$.
Boundary conditions are incorporated using a ghost-point approach. These conditions enter the discrete differential operators as additive correction terms that augment the interior FD stencil. Thereby, each operator application can be expressed as a matrix–vector product plus an explicit boundary-correction vector, while boundary points themselves are excluded from the solution vector~\cite{Ferziger2020}.
Contributions to the source term $\vecc{f}$ as volume-specific body forces by the physical problem are neglected in the course of this work.

\subsection{Discretization and Representation}
\label{sec:disc}
The spatial discretization of the two-dimensional physical domain $\Omega$ is based on a structured O-type grid $\mathcal{G}$ \cite{Thompson1974, Thompson1982}. As outlined in Appendix~\ref{App:DiscretizationCurvis}, the physical coordinates $(x_{ij}, z_{ij})$ are defined by a mapping to a unit-square computational domain $\Omega_0$ using a simple transfinite interpolation approach \cite{Gordon1973}.
The interior supporting points of the computational domain are located at
\begin{equation}
   \label{eq:defetaxi}
    \begin{rcases2}
  \eta_{ij}&=(i+1) \delta \eta \  \\
  \xi_{ij}&=j \delta \xi \ 
    \end{rcases2} \, 
\text{with } 
    \begin{array}{l}
        i = 0, \ldots, 2^{n_\eta} - 1\ , \\
        j = 0, \ldots, 2^{n_\xi} - 1 \ . 
    \end{array}
\end{equation}
Hence, employing \(n_\eta\) and \(n_\xi\) TT-cores for the \(\eta\)- and \(\xi\)-directions, respectively. We set \(N_\eta = 2^{n_\eta}\) and \(N_\xi = 2^{n_\xi}\), thereby obtaining an \(N_\eta \times N_\xi\) structured grid with uniform spacings
\(\delta\eta = 1/(N_\eta + 1)\) and \(\delta\xi = 1/N_\xi\) (cf.\ Appendix~\ref{App:DiscretizationCurvis}).
Note that the auxiliary indices \(i=-1\) and \(i=N_\eta\) coincide with the
inner (top) and exterior (bottom) boundaries of $\Omega$ ($\Omega_0$) domain, while the \(\xi\)-direction is
treated periodically.

Next, we define a coordinate transformation $\Phi: {\Omega_0} \mapsto \Omega$ that maps the computational coordinates $(\xi_{ij}, \eta_{ij})$ to the physical coordinates $(x_{ij}, z_{ij})$ with $x_{ij}=x(\xi_{ij}, \eta_{ij})$ and $z_{ij}=z(\xi_{ij}, \eta_{ij})$~\cite{Thompson1982}.
Using this transformation, we approximate the spatial derivatives in the momentum Eqs.~\eqref{eq:Navier-Stokes-Eq-B} and continuity Eq.~\eqref{eq:Conti-Eq-B} by their analogues in the computational domain, thereby obtaining curvilinear forms of the differential operators.

\subsubsection{2D Field Encoding}\label{subsub:QTT:fluid_field_enc}
\begin{figure}[t]
    \centering
    \includegraphics[width=0.9\linewidth]{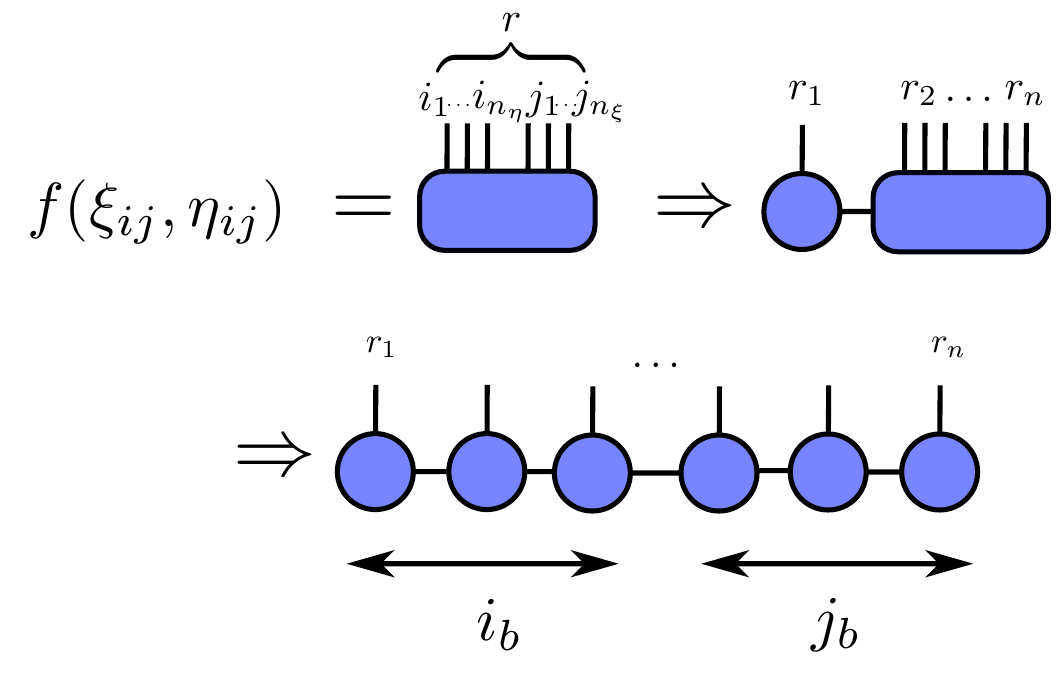}
    \caption{Graphical representation of a TT decomposition of a 2D field.
    Here, $i_k$ ($j_k$) represents the value of the $k$\textsuperscript{th} bit of bit string $i_b$ ($j_b$).}
    \label{fig:2-index-QTT-decomp}
\end{figure}

Before proceeding with the discretization in the TT format, we first introduce the encoding of two-dimensional scalar fields, here defined over the computational domain $\Omega_0$, as TT-vectors. Consider, for example, a scalar field $a(\xi_{ij}, \eta_{ij})$, such as the $x$-coordinate $x(\xi_{ij}, \eta_{ij})$. To represent such a field as a TT-vector, a consistent enumeration of the grid points $(\xi_{ij}, \eta_{ij})$ must first be established.
To each point $(\xi_{ij}, \eta_{ij})$, we assign a bit string $\physone{} = i_b\cup j_b $ of length $n = n_{\eta} + n_{\xi}$, formed by concatenating $i_b$, the length $n_\eta$ binary representation of row index $i\in\{0,1,\ldots N_\eta-1\}$, with $j_b$, the length $n_{\xi}$ binary representation of column index $j\in\{0,1,\ldots N_\xi-1\}$. Note that $i$ and $j$ are the indices defined in Eq.~\eqref{eq:defetaxi}. 
Each bit in $\physone{}$ is then interpreted as the two modes $\{0,1\}$ of the indices of an 
order-$n$ tensor $\tensor{a}_{\physone{1},\physone{2},\ldots,\physone{n}}$ that encodes $N=2^{n}$ discretized field values
as presented in Section~\ref{sec:QTT_Principles}, where $\physone{k}$ is the $k$\textsuperscript{th} bit of $\physone{}$. 
After decomposing $\tensor{a}_{\physone{1},\physone{2},\ldots,\physone{n}}$ for $n-1$ times, 
as illustrated in Fig.~\ref{fig:2-index-QTT-decomp}, one obtains the TT-vector representation of the field
\begin{equation}
    \qtt{a} =
   \tensor{A}[1]^{\physone{1}}\tensor{A}[2]^{\physone{2}} \ldots \tensor{A}[n]^{\physone{n}}
   \simeq \tensor{a}_{\physone{1},\physone{2},\ldots,\physone{n}} 
= f(\xi_{ij},\eta_{ij})\, .
\end{equation}

Vector fields, such as the velocity $\vecc{v}$, are treated by encoding $x$- and $z$-components of a field separately into two different TT-vectors.

While the aforementioned method yields a flexible encoding of various fields, the computational cost is proportional to the number of grid points $N=N_{\eta}N_{\xi}$. 
Practically, a field with recognized features (e.g.,~uniform, harmonic, exponential, ...) has a low-rank representation that allows a direct construction of its TT-cores that only costs $\mathcal{O}\big(\log{(N)}\chi^2\big)$~\cite{Oseledets2012}, or it can be found through tensor cross interpolation~\cite{Ritter24}.
For the simulations conducted in this paper, the TT-vectors of the initial flow fields are directly constructed within the TT format, thus avoiding the decomposition of order-$n$ tensors. By keeping the bond dimension fixed to a constant value throughout each step of the algorithm described in Section~\ref{sec:algorithm}, the time evolution and subsequent analysis are performed without requiring the full numerical representation of the fields, thereby preserving the overall logarithmic scaling in $N$.

\subsubsection{Mesh Generation in TT Format}\label{sec:QTT:mesh_gen}
The discretization $( x_{ij},z_{ij} )$ is classically defined by a grid $\mathcal{G}$, which can either be read in or generated directly in the TT format.
In either case, the number of interior grid points must satisfy the constraint $2^{n_\eta} \times 2^{n_\xi}$ to be compatible with the TT format. 
To preserve the logarithmic scaling in the number of points $N$, the latter approach is preferable, as it avoids the explicit construction and decomposition of full $N$-component tensors, and is hence discussed in the following.

For a convex shape $\Gamma$ with supporting points on the surface given by
$(x_{\Gamma \, j}, z_{\Gamma \, j})$, the interior grid points can be constructed either uniformly along the surface normal direction, 
or using a non-uniform distribution with step sizes that gradually increase with the distance from the surface $\Gamma$, e.g.,
\begin{align}
    s_i^{_{(\text{lin})}} &= Q (i+1) \delta \eta \quad \quad \ \ \ \text{(linear)} \quad \text{or}\\
    s_i^{_{(\text{exp})}} &= Q \frac{e^{\kappa (i+1) \delta \eta }-1}{e^{\kappa}-1}  \quad \text{(exponential)}\, ,
\end{align}
where $Q$ denotes the maximal normal distance to the surface $\Gamma$ and $\kappa$ is a stretching parameter. 
For ${\kappa \rightarrow 0}$, the exponential $s_i^{_{(\text{exp})}}$ spacing reduces to linear $s_i^{_{(\text{lin})}}$.
The grid coordinates are constructed as a tensor product of 1D functions
defined along the $\eta$- and $\xi$-directions, respectively:
\begin{align}
     x_{ij} &= 1_i x_{\Gamma \, j} + s_i \, \nx \label{eq:linear_mesh} \quad \text{and}\\
     z_{ij} &= 1_i z_{\Gamma \, j} + s_i \, \nz \, , \label{eq:exp_mesh}
\end{align}
where $s_i$ is either $s_i^{_{(\text{lin})}}$ or $s_i^{_{(\text{exp})}}$, $1_i$ denotes a vector of ones, and
$\nx$, $\nz$ are the $x$- and $z$-components of the unit normal vector at surface point $j$. For the orthogonal cylinder mesh illustrated in Fig.~\ref{fig:cylinder}, Eqs.~\eqref{eq:linear_mesh}, \eqref{eq:exp_mesh} agree with the results of a slightly more involved classical transfinite interpolation method~\cite{TFI1973}.

To translate Eqs.~\eqref{eq:linear_mesh} and \eqref{eq:exp_mesh} into TTs, the $N_\xi$-component vectors containing to the $x$- and $z$-coordinates of the object's surface $\Gamma$, $x_j^{_{( \Gamma)}}$ and $z_j^{_{( \Gamma)}}$, as well as the components of the unit normal vector, $\nx$ and $\nz$, with $j=0,\ldots,N_\xi-1$
must each be encoded as TT-vectors of length $n_\xi$. These are denoted as $\qtt{x}^{( \Gamma)}$, $\qtt{z}^{( \Gamma)}$, \( \nxqtt \), and \( \nzqtt \), respectively.
Furthermore, the TT-vectors of length $n_\eta$ corresponding to $1_i$ and $s_i$, with $i = 0, \ldots, N_\eta - 1$, are analytically prescribed and low-rank. 
The TT-cores of $1_i$ and linear spacing $s_i^{(\text{lin})}$ can be constructed following Ref.~\cite{Oseledets2012}, while the strategy for the exponential spacing $s_i^{(\text{exp})}$ is described in Ref.~\cite{Siegl2025}, and denoting them by $\qtt{1}$ and $\qtt{s}$. 
The final TT-vectors of the interior coordinates ($x_{ij} \rightarrow \qtt{x}$ and $z_{ij} \rightarrow \qtt{z}$)  are then realized by concatenating the above TT-vectors as
\begin{align}
    \qtt{x} &= \qtt{1} \cup \qtt{x}^{( \Gamma)} + \qtt{s}  \cup \nxqtt \quad \text{and}\\
    \qtt{z} &= \qtt{1} \cup \qtt{z}^{( \Gamma)} + \qtt{s}  \cup \nzqtt \, ,
\end{align}
where $\cup$ here denotes $\qtt{a} \cup \qtt{b} = \tensor{A}[1] \ldots \tensor{A}[n_\eta]\,\tensor{B}[1]\ldots \tensor{B}[n_\xi]$. The resulting TT-vector is then of length $n_\eta+n_\xi$.

\subsubsection{Constructing Curvilinear Operators in TT Format}
\label{subsub:QTT:Curvis}

The construction of curvilinear operators as TT-matrices exemplarily presented for the physical-space derivative $\partial_x $ follows (cf.~\hyperref[App:DiscretizationCurvis]{Appendix~\ref{App:DiscretizationCurvis}})
\begin{equation}
     \partial_x = \frac{1}{J} \Big((\partial_\eta z) \partial_\xi - (\partial_\xi z)  \partial_\eta \Big)\, , 
\end{equation}
where $J$ is the Jacobian determinant $J =  (\partial_\xi x) (\partial_\eta z) - (\partial_\eta x) (\partial_\xi z) $. 
The same construction strategy applies analogously to the physical-space derivative $\partial_z$ as well as to the Laplacian operator $\Delta$. We note that all computational-space derivatives are discretized using a second-order central FD scheme (cf.~Appendix~\ref{App:DiscretizationCurvis}).

Given the TT-vector representations of the internal grid points, $\qtt{x}$ and $\qtt{z}$, as detailed in Section~\ref{sec:QTT:mesh_gen}, the derivative operators $\partial_\xi$ and $\partial_\eta$ can be directly constructed as low-rank TT-matrices~\cite{Kazeev2012}, $\qtt{D}_\xi$ and $\qtt{D}_\eta$, respectively.
Accordingly, the TT-vector $\qtt{j}$ of the Jacobian determinant $J$ is computed as  
\begin{equation}
          \qtt{j} =    (\qtt{D}_\xi \qtt{x}) (\qtt{D}_\eta \qtt{z}) - (\qtt{D}_\eta \qtt{x}) (\qtt{D}_\xi \qtt{z})\, .
\end{equation}
Next, the computation of
${\qtt{c}_1 = (\qtt{D}_\eta \qtt{z}) \odot \qtt{j}^{-1}}$ and ${\qtt{c}_2 = (\qtt{D}_\xi \qtt{z}) \odot \qtt{j}^{-1}}$ is executed where the inverse $\qtt{j}^{-1}$ is found as the solution of the LSE $\qtt{J} a = \qtt{1}$,
where $\qtt{J}$ denotes the diagonal TT-matrix representation of $\qtt{j}$ and \qtt{1} a TT-vector of ones. Thus $\qtt{c}_1$ and $\qtt{c}_2$ are found by a combination of known operations, cf.~Section~\ref{sec:QTT:Elementwise}-\ref{sec:QTT:SLE}.
As a final step, the prefactors $\qtt{c}_1$, $\qtt{c}_2$ represented as TT-vectors, need to be multiplied into the TT-matrices $\qtt{D}_\xi$ and $\qtt{D}_\eta$ to construct the partial derivative in $x$-direction in TT format, viz. $\qtt{D}_x = \qtt{c}_1  \qtt{D}_\xi -  \qtt{c}_2  \qtt{D}_\eta$. This operation is realized as
\begin{equation}
    \qtt{D}_x = \qtt{C_1} \qtt{D}_\xi -  \qtt{C}_2 \qtt{D}_\eta\, ,
\end{equation}
where $\qtt{C}_1$ ($\qtt{C}_2$) are diagonal TT-matrices
that have $\qtt{c}_1$ ($\qtt{c}_2$) on its diagonal using the $\pmb{\delta}^{\physone{k}'}_{\physone{k},\physone{k}''}$ tensor
as introduced in Section~\ref{sec:QTT:Elementwise}.
The matrix-matrix product $\qtt{C_1} \qtt{D}_\xi$ is performed similarly to a TT-matrix with TT-vector contraction 
with changes w.r.t.~the physical dimension, and, in the worst case, scaling as $\mathcal{O}\big(\chi(\qtt{D}_{\xi})^3 \chi(\qtt{C}_1)^3\big)$).
Note that the bond dimension $\chi(\qtt{C}_1)$ generally depends on the grid coordinates, while $\chi(\qtt{D}_\xi)$ is low-rank and independent of the grid~\cite{Kazeev2012}. Moreover, they are independent of the flow regime, as it requires the construction of the curvilinear operators only once for a given FD scheme and discretization.
Lastly, in constructing the curvilinear operators as TT-operators, a truncation is performed after each TT operation by discarding singular values below $10^{-14}$ (cf.~Section~\ref{sec:QTT_Principles}).

\subsection{Fractional Step Method}
\label{sec:algorithm}
A fractional step method~\cite{Ferziger2020} is used to advance the flow field described by the momentum Eqs.~\eqref{eq:Navier-Stokes-Eq-B} and the continuity Eq.~\eqref{eq:Conti-Eq-B} from $t^{(m)}$ to $t^{(m+1)}=t^{(m)}+\delta t$. 
In this approach, the pressure gradient-free momentum equations are used to determine an intermediate velocity $\vecc{v}^*$ \cite{Ascher1997}
 \begin{equation}
    \begin{split}
        &\frac{\vecc{v}^*-\vecc{v}^{(m)}}{\delta t} + \big(\vecc{v}^{(m)} \cdot \vecc{\nabla} \big)\vecc{v}^{(m)} = \frac{1}{Re} 
        \Delta \, \vecc{v}^*
        \\
        &\Leftrightarrow \left(1 - \frac{\delta t}{Re} 
        \Delta \right) \vecc{v}^* = \vecc{v}^{(m)} - \delta t \big( \vecc{v}^{(m)} \cdot \vecc{\nabla} \big) \vecc{v}^{(m)} 
        .  \label{eq:intermediate_vel_eq}
    \end{split}
    \end{equation}
The intermediate velocity $\vecc{v}^*$ is subsequently corrected by solving a pressure Poisson equation~\cite{Ferziger2020}, which is controlled by the divergence of the intermediate velocity field, i.e.,
  \begin{equation}
        \Delta \, p^{\vecc{}(m+1)} =  \frac{1}{\delta t} \, \vecc{\nabla} \cdot \vecc{v}^*
        \,. \label{eq:Poisson}
    \end{equation}
The gradient of the computed pressure $p^{(m+1)}$ is used to enforce the continuity constraint for the velocity at the next time step $t^{(m+1)}$ with 
\begin{equation}
        \vecc{v}^{(m+1)} = \vecc{v}^*-\delta t \vecc{\nabla} \, p^{(m+1)}\,.
\end{equation}

Instead of an adaptive, Courant-based time step, $\delta t$ is kept constant to maintain the efficiency in the TT format. The resulting method is equivalent to the fractional step method originally proposed by~\citeauthor{Chorin1968}~\cite{Chorin1968}.
The Poisson equation~\eqref{eq:Poisson} is solved in TT format using the variational optimization outlined in Section~\ref{sec:QTT:SLE}.
This procedure is particularly well suited to elliptic equations, and we include the Jacobian determinant~$J$ to account for the varying cell sizes in the curvilinear discretization~\cite{Grossmann2007}.
Consequently, the modified optimization problem for the Poisson Eq.~\eqref{eq:Poisson} reads
\begin{equation}
\min \limits_{\qtt{a}} \mathcal{C}(\qtt{a}):=  \frac{1}{2} \left(\qtt{\qtt{a}} \left(\qtt{j}\qtt{M}^{(\Delta)}\right) \qtt{\qtt{a}} \right) -  \qtt{a} (\qtt{j} \odot \qtt{b}) \,. 
\end{equation}

To facilitate the direct comparison of the TT solver with its classical counterpart, the fractional step method described above is implemented twofold. The first implementation follows the TT methodology, while the second refers to a classical FD implementation, which is verified against literature and finite-volume results generated with OpenFOAM. The interested reader finds more information in Appendix~\ref{app:openfoam}.

\section{Applications}
\label{sec:results}
In this section, we present a detailed comparison between our TT solver and a reference classical FD solver for the paradigmatic 2D cylinder flow, considering both a non-rotating and a rotating cylinder. We start by introducing the specific problem setup of the cylinder and its discretization. Further we introduce flow characteristics and diagnostic quantities used to evaluate solution accuracy and computational efficiency.
Results of the first Subsection~\ref{sec:cylinder} are dedicated to the non-rotating cylinder in steady ($Re=20$) and transient ($Re\geq50$) regimes. The subsequent Subsection~\ref{sec:rot_cylinder} discusses the transient flow around the rotating cylinder ($Re=100$). Section~\ref{sec:Performance} complements the numerical results with insights into computational
performance. All runs were carried out on $8$ cores of an AMD EPYC 9654 CPU and $32$ GB of RAM, except stated otherwise.

\begin{figure}[htbp]
    \centering
         \def\svgwidth{0.45\textwidth}
         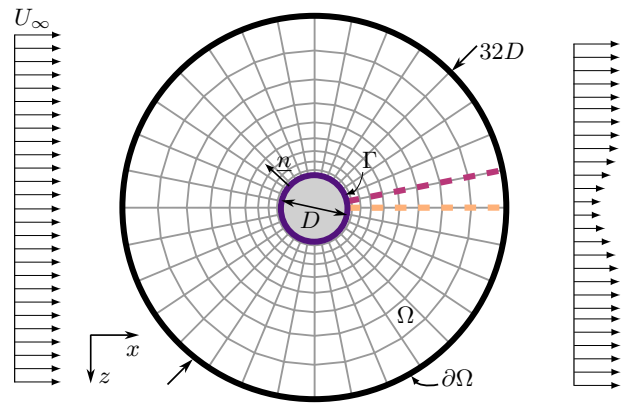  
         \caption{Dimensionless illustration of the investigated, discretized 2D cylinder case
         with the grid lines indicated in light-gray and the exterior (interior) boundary of the physical domain $\Omega$ indicated in black (magenta) (cf.~Appendix~\ref{App:DiscretizationCurvis}).}
    \label{fig:cylinder}
\end{figure}

\smallskip 
Figure~\ref{fig:cylinder} depicts the test case for a cylinder that features a unit diameter $D$ boundary $\Gamma$ which is centered 
in a circular domain $\Omega$ with the boundary $\partial\Omega$ placed with diameter $32\dcyl$.
The spatial discretization refers to a block-structured O-type grid, as shown by the light-gray grid lines in Fig.~\ref{fig:cylinder}. Three successively refined grids, each equidistant in the circumferential direction, are considered:
\begin{equation}
    \begin{split}
        \mathcal{G}_1 &: N = 256 \times 256,\ \text{\nvector{8}{8}}\ ,\\
        \mathcal{G}_2 &: N = 256 \times 512,\ \text{\nvector{8}{9}}\ ,\\
        \mathcal{G}_3 &: N = 512 \times 512,\ \text{\nvector{9}{9}}\ ,
    \end{split}
\end{equation}
with minimal grid spacings at the cylinder surface, each taken as the smaller of the normal spacing or the circumferential arc length, that are approximately ${h_{\min} \approx 6.78\times10^{-3}}$, ${h_{\min} \approx 6.14\times10^{-3}}$ and ${h_{\min} \approx 3.39\times10^{-3}}$, respectively. Here, ${h_{\min} = \tilde h_{\min}/D}$ denotes the smallest step size in the physical domain which usually occurs at the interior boundary in the radial direction. The grid points are specified in Appendix~\ref{app:gridpoints}.
While the grids can be constructed directly within the TT format as outlined in Section~\ref{sec:QTT:mesh_gen}, in this study we construct the grid points ($x_{ij}, z_{ij}$) and the associated metric terms in a classical way and subsequently encode them into TT-vectors. This ensures comparability between FD and TT-based solvers, which therefore operate with identical coordinate transformations. The computational overhead due to the transformation of the classical grid into the TT format is neglected here. Yet, its advantageous efficiency should be exploited for any other application.

The assessment of transient flow phenomena employs the Strouhal number 
\begin{equation}
St= {\tilde\omega} \dcyl /2\pi U_\infty \label{eq:st} \, ,
\end{equation}
where ${\tilde\omega}\,[\qty{}{rad \per s}]$ is the circular frequency of the oscillating lift force $\tilde F_\text{L}$.
The lift force $\tilde F_\text{L}$ is defined as the force per unit length acting perpendicular to the approach flow, whereas the drag force $\tilde F_\text{D}$ acts parallel to it.
The corresponding non-dimensional coefficients read
\begin{equation}
\label{eq:liftdrag}
    C_\text{L} =2 \tilde F_\text{L}/(\tilde{\rho} U_\infty^2 \, D) \quad \text{and} \quad C_\text{D}=2 \tilde F_\text{D}/(\tilde{\rho} U_\infty^2 \, D) \, .
\end{equation}

The predictive accuracy is assessed by the $L_2$-norm of a scalar field $s(\xi, \eta) $, which accounts for the variable grid spacing across the domain, i.e.,
\begin{equation}
    \begin{split}
    ||s||_{\text{L}_2} &= \sqrt{\int_{\Omega_0} |s(\xi, \eta)|^2 J d\xi d\eta } 
    \approx \sqrt{\delta \xi \delta \eta
    |s_{ij}|^2 J_{ij}}\, .
    \end{split}
\end{equation}
Consequently, a relative difference between a TT solution $\qtt{s}$ and a classical solution $s$ reads
\begin{equation}
    \epsilon_{\text{L}_2} = \frac{||\qtt{s}-s||_{\text{L}_2}}{||s||_{\text{L}_2}}\, ,
    \label{eq:error}
\end{equation}
and is used to evaluate the error of the TT format. We also consider the maximal difference error
\begin{equation}
    \epsilon_{\max} =  \|s - \qtt{s}\|_\infty.
    \label{eq:max_error}
\end{equation}
Moreover, time averages are denoted by an overline, e.g.,~the error $\overline{\epsilon_{\text{L}_2}}$.
To quantify the compression achievable through the TT representation, we introduce the number of variables parameterizing the solution (NVPS), which is the total number of values needed to be stored to represent a state~\cite{Kiffner2023}. The NVPS of a TT-vector depends on the dimensions of its indices and is computed by
\begin{equation}
       \text{NVPS}_{\text{TT}} = 2 \sum_{k=1}^{n} \linkone{k-1}  \linkone{k} \, ,
\end{equation}
where \(\linkone{k}\) denotes again the bond index between the \(k\)-th and \((k+1)\)-th TT-core, and $n=\log{(N)}$ the number of TT-cores. Therefore, the NVPS for a TT-vector grows only as $\mathcal{O}(n \chi^2)$~\cite{Kiffner2023}. In contrast, in a classical FD description, the solution vector length grows exponentially as $\mathcal{O}(2^n)$.s

\subsection{Non-Rotating Cylinder}
\label{sec:cylinder}
Within this section, the non-rotating cylinder is modeled by zero velocity conditions such that $\vecc{v}_\Gamma=0$ in Eq.~\eqref{eq:bound}. 
The initial Section~\ref{subsec:Steady} compares the INSE solutions (\ref{eq:Navier-Stokes-Eq-B}-\ref{eq:bound}) obtained from the classical FD method with TT-based solutions for the steady state flow at $Re=20$, and scrutinizes the accuracy of the TT's steady-state solution for several bond dimensions. The subsequent Section~\ref{subsec:unsteady} investigates the transient regime for increasing Reynolds numbers, where emphasis is given to the ability of TT to accurately predict vortex shedding and the related oscillating forces. 

\subsubsection{Steady Flow (Re=20)}
\label{subsec:Steady}
To study the effects introduced by the compression inherent to the TT format, the differences between both strategies are analyzed for increasing bond dimensions~$\chi$. We note that the bond dimension is imposed uniformly on the TT-vectors of the two cartesian velocities $ u $, $ v $, and the pressure $ p $, such that each field has the same value of~$\chi$. For the sake of completeness, the accuracy of the classical FD code is verified against a frequently used finite volume procedure in Appendix~\ref{app:openfoam}.
For the stability criterion of the convective term, a maximum velocity as $U_{\max} =\gamma U_\infty$ is estimated following the usual practice in the literature
(cf.~\cite{Ferziger2020}); here we adopt \(\gamma = 1.5\). 
Accordingly, the simulation on the medium–fine grid
\(\mathcal{G}2\) is advanced with a time step $
\delta t = \tilde{\delta t}\,D /U_\infty\approx 3.8\times10^{-3}$,
so that the resulting convective Courant number remains below unity.

\begin{figure*}[htbp]
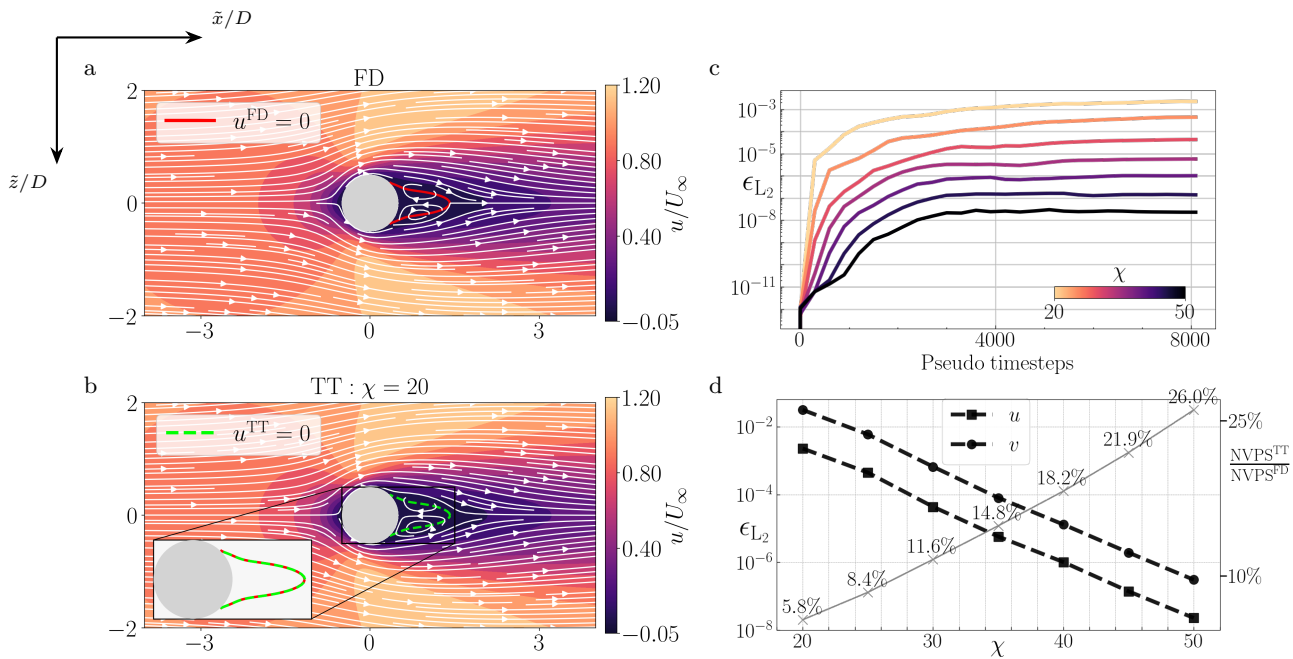

  \centering
    \include{Figures/SteadyRe20/fig_8_tikz_code}
  \caption{
  Streamlines and horizontal velocity ($u$) contours for the steady‐state flow at $Re=20$, predicted by (a) the classical FD method and (b) the TT method with $\chi=20$. The $u=0$ isoline is highlighted in solid red for FD and dashed green for TT, with an inset in panel (b) showing a close-up comparison of both isolines.
  Predictive difference $\epsilon_{\text{L}_2}$ (\ref{eq:error}) between the velocity magnitude field
  (\hyperref[fig:res-figure-1]{c}) and the individual Cartesian velocity components $u$, $v$ (\hyperref[fig:res-figure-1]{d}) obtained from the TT and the classical FD simulations for the flow around the cylinder at $Re=20$ using the medium-fine grid $\mathcal{G}2$.
  Panel~\hyperref[fig:res-figure-1]{(c)} shows the temporal evolution for different bond dimensions $\chi$.
  Panel~\hyperref[fig:res-figure-1]{(d)} depicts the steady state results for $u$ (circles) and $v$ (squares), supplemented
  by the NVPS fraction for each $\chi$ (crosses) to indicate the corresponding compression rate.}
  \label{fig:res-figure-1}
\end{figure*}
Starting from a uniform initial condition with $\vecc{v}_\text{init.}=(U_\infty,0)^\intercal$, solutions are extracted after the system evolves over $8100$ steps in pseudo time, which corresponds to more than 30 passage times $D/U_\infty$. 
Figure~\ref{fig:res-figure-1}(a) displays the steady-state streamlines and horizontal-velocity ($u$) contours at $Re = 20$ predicted by the classical FD method, while Fig.~\ref{fig:res-figure-1}(b) shows the corresponding TT solution with $\chi = 20$. In both panels the $u = 0$ isoline is depicted, solid red for the FD and dashed green for the TT. Moreover, Fig.~\ref{fig:res-figure-1}(b) includes an inset that magnifies these isolines for direct comparison.
Qualitatively, the predicted fields show excellent agreement, and the classical F\"oppl vortex pair behind the cylinder is correctly captured.

Figure~\hyperref[fig:res-figure-1]{\ref*{fig:res-figure-1}\,(c)} depicts the evolution of the relative difference (error) $\epsilon_{\text{L}_2}$ between the velocity magnitude obtained from TT-based simulations and FD predictions, evaluated across a range of fixed bond dimensions $\chi$. 
The error increases during the initial transient phase before stabilizing after around 25 passage times at distinct levels for each  $\chi$, remaining below $2\times10^{-3}$.
The maximum difference in $u$ is bounded by $\epsilon_{\max}\le 9.9\times10^{-3}$ ($\chi=20$), decreasing to $2.2\times10^{-7}$ ($\chi=50$). Similarly, the $v$-field satisfies $\epsilon_{\max}\le 1.3\times10^{-2}$ ($\chi=20$), dropping to $\epsilon_{\max}\le 7.9\times10^{-8}$ ($\chi=50$).
The negligible lift coefficient of $\lvert C_{\mathrm{L}} \rvert = 1\times10^{-4}$ for the FD solver exhibits its largest deviation w.r.t.~the TT solution at $\chi = 25$ with $\lvert C_{\mathrm{L}} \rvert = 3\times10^{-3}$. 
For all the bond dimension $\chi \ge 30$ the lift coefficient is upper bounded by $\lvert C_{\mathrm{L}}\rvert \le 1.7\times10^{-4}$, closely resembling the FD solution.
Moreover, in the energy norm commonly used to assess ROMs (cf.~Ref.~\cite{Koc2025}), the steady-state relative error in total kinetic energy $E_{\text{kin}}=1/2 \, \|\vecc{v}\|_{\text{L}_2}^2$ compared to FD decays rapidly with bond dimension, from $7.65\times10^{-5}$ at $\chi=20$, to $7.16\times10^{-8}$ at $\chi=35$, and down to $3.62\times10^{-11}$ at $\chi=50$.
Figure~\hyperref[fig:res-figure-1]{\ref*{fig:res-figure-1}\,(d)} shows the final differences in $u$ and $v$ after $8100$ time steps.

In general, a remarkable agreement between the TT and FD solution is observed. 
It is evident that with approximately $5.8\%$ of the NVPS, the TT-based approach achieves a relative error of $3 \times 10^{-3}$ (or $0.3 \%$)
in the velocity magnitude,
achieving an outstanding $20$-fold compression. Increasing the NVPS fraction to approximately $15\%$ further reduces the relative error by three magnitudes to $5.8 \times 10^{-6}$.
In the most demanding case with $\chi=50$, the method still requires only around $1/4$ of the NVPS while achieving highly accurate solutions with relative errors on the order of $10^{-8}$. An extrapolation of the errors suggests that the accuracy can in principle be reduced to machine precision by further increasing the bond dimension.
More details will be found in the analysis of the time-dependent flow in the following section.

\subsubsection{Transient Flow}
\label{subsec:unsteady}
We now proceed to analyze unsteady flow regimes that emerge for $Re\geq50$.
Our goal is to demonstrate that TT can not only efficiently capture simple steady-state flows but also faithfully reproduce transient dynamics. In particular, we investigate the trade-off between parametrization (NVPS) complexity and pointwise solution errors, noting that exact local accuracy may not be required to capture the essential dynamics, aligning with the core idea of ROMs.
The simulations apply the two coarser grids $\mathcal{G}1$ and $\mathcal{G}2$ with a time step of $\delta t  =\tilde{\delta t} U_\infty/D  \approx 3.4 \times 10^{-3} $ and $ \delta t =\tilde{\delta t} U_\infty/D \approx 3.1 \times 10^{-3} $, respectively. Here, the maximum velocity is estimated with $\gamma=2$, i.e.,~$U_{\max}= \gamma U_\infty$, so that the resulting Courant number remains below unity for all $Re$ considered. Expecting a Strouhal number of $St \approx 0.2$ corresponds to a temporal resolution of approximately 60 time steps per period.
Two algorithmic modifications are employed to compute the transient flows: (1)
we trigger transient effects by employing asymmetries within initial conditions. To this end, we add an initial perturbation $u_{\text{pert.}}$ to the initial velocity field, which consists of a 5\% velocity modulation (as given in Appendix~\ref{app:AlgorithmModificationsforTransient}) and (2) we apply a preconditioning of the Poisson equation within the TT format, allowing to efficiently simulate higher bond dimensions~\cite{Dolgov2012}.

\begin{figure}[t]
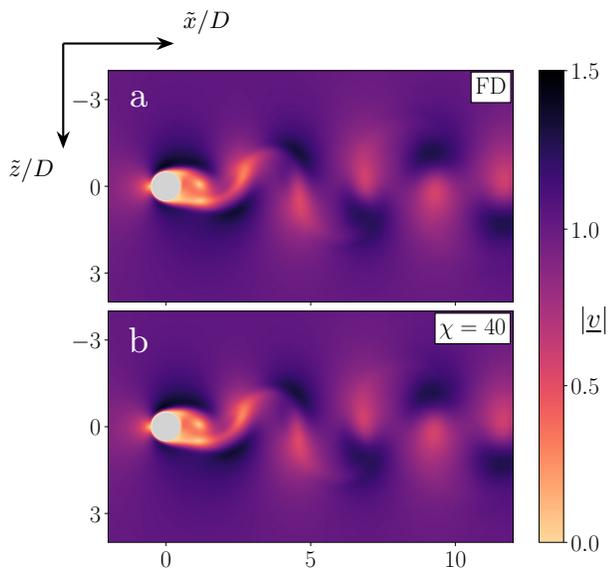

\centering
\include{Figures/TransientSec/fig_9_tikz_code}
    \caption{Snapshots of velocity magnitude $|\vecc{v}|$ extracted at $t D/U_\infty =84.85$ from TT in (a) ($\chi=40$) and a classical FD (b) simulation at $Re=150$ when using the coarse grid $\mathcal{G}1$. 
    }

    \label{fig:Snapshots_Karman}
\end{figure}

\begin{figure}[t]
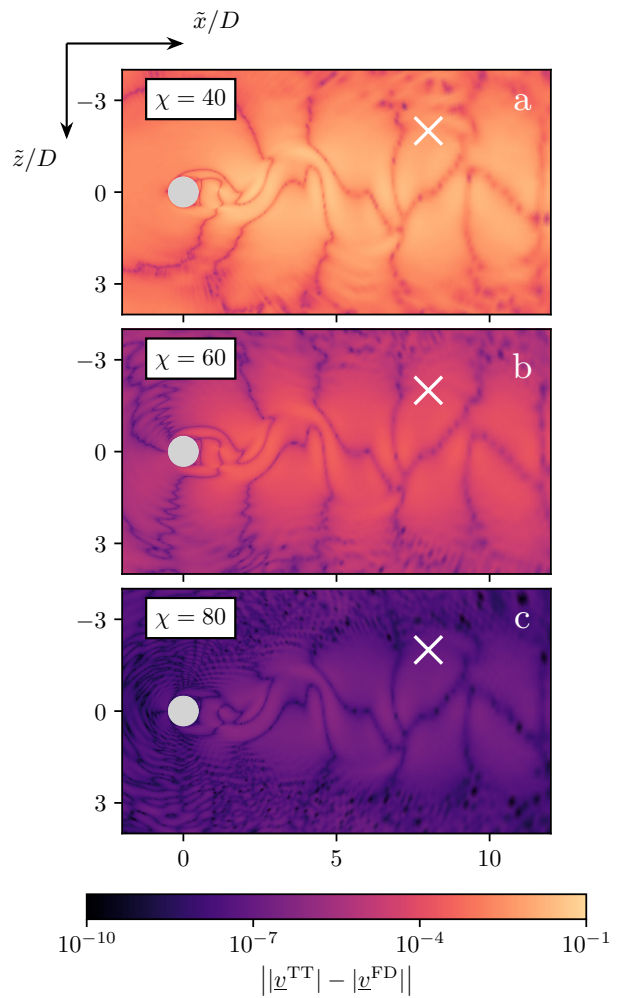

  \centering
\include{Figures/TransientSec/fig_10_tikz_code}
    \caption{
   Pointwise difference of the velocity magnitude $|\vecc{v}|$  predicted by the TT and classical FD method for the flow around a non-rotating cylinder at $Re=150$ on the coarse grid $\mathcal{G}1$.
   White crosses indicate the probe location $(\tilde{x}/D, \tilde{z}/D) = (8, 2)$, for the comparison of velocity components.
   }
   \label{fig:snapshots_error_distr}
\end{figure}

Figure~\ref{fig:Snapshots_Karman} displays snapshots of the TT for $\chi=40$ and the FD-predicted velocity magnitudes $|\vecc{v}|$ for $Re=150$. Results refer to the coarse grid $\mathcal{G}1$ at around $84.85$ passage times, i.e.,~approximately $2.5 \times 10^4$ time steps when the flow has reached a periodic state.
Qualitatively, both velocity fields are almost indistinguishable from each other, demonstrating that the TT simulation at $\chi=40$ is capable of capturing the K\'arm\'an vortex street similar to the classical approach.
Figure~\ref{fig:snapshots_error_distr} illustrates the corresponding pointwise differences of the velocity magnitude $|\vecc{v}|$, defined as $\left| |\vecc{v}^{\text{TT}}| - |\vecc{v}^{\text{FD}}| \right|$, in the region $\tilde{x}/D\in[-2,12], \tilde{z}/D\in [-4,4]$.
The bond dimension $\chi$ increases from top to bottom across the panels, with $\chi=40, \, 60$, and $80$. Results are again  evaluated after around $2.5\times10^4$ time steps, when the flow reflects a periodic behavior in time. 

Figure~\ref{fig:snapshots_error_distr} demonstrates that each increase of $\chi$ by $20$ bond units reduces the local deviation by roughly two orders of magnitude.
For instance, at the location $(\tilde{x}/D, \tilde{z}/D) = (8, 2)$ (marked by '$\times$' in Fig.~$\ref{fig:snapshots_error_distr}$), the deviation reads $5.2 \times 10^{-3}$, $6.3 \times 10^{-5}$, and $9.7 \times 10^{-8}$ for increasing values of $\chi$. This aligns with the convergence of TT towards FD in the limit of increasing $\chi$.
For the given system, the maximum possible bond dimension is $\chi=256$. 
With a reduction to $\chi=80$ and a corresponding $87\%$ of NVPS compared to FD, a maximum error of $2.54 \times 10^{-6}$ in the velocity magnitude $|\vecc{v}|$ is recovered, cf.~\hyperref[fig:mpo_coeffs]{Fig.~\ref{fig:snapshots_error_distr}\,(c)}.
Nevertheless, lower $\chi$ values are sufficient to capture the essential physics for this $Re$, as it was concluded before.
This is confirmed by the error patterns in the cylinder's wake.
As expected, the error patterns are aligned to the wake and appear similar for different values of $\chi$. This indicates a consistent resolution of the dynamics with only differences in the resolution accuracy, and underlines the suitability of the TT-based approach as a ROM.

\begin{figure}[htbp]
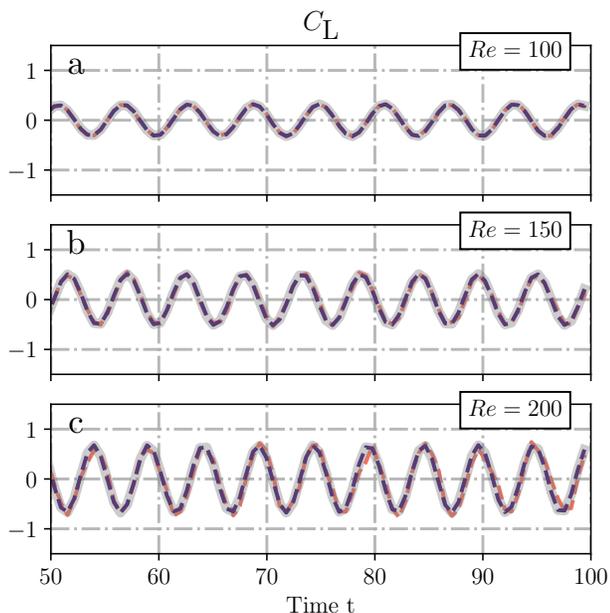

    \centering
    \include{Figures/TransientSec/fig_11_tikz_code}
    \caption{
    Temporal evolution of the lift coefficient $C_\mathrm{L}$ for the flow around a non-rotating cylinder on the medium grid $\mathcal{G}_2$.  
    Results predicted by the TT method are shown for bond dimensions \(\chi=40\) (orange dashed) and \(\chi=60\) (magenta dashed), with the FD reference solution (gray solid).
    }
    \label{fig:lift_oscillation}
\end{figure}

The capability of TTs to predict the relevant integral  force quantities is assessed next. To this end, the Strouhal number is determined from the oscillations of the lift force coefficient $C_{\text{L}}$ (\ref{eq:liftdrag}).
Accordingly, the computation of $C_{\text{L}}$ can be performed entirely in the TT format, without requiring an expensive contraction to the full classical representation. 
Specifically, it involves extracting one-dimensional TT-vectors of $\eta$-constant lines adjacent to the surface $\Gamma$ (cf.~Ref.~\cite{Kiffner2023}), combining TT operations outlined in Section~\ref{sec:QTT_Principles}, and applying a summation (average-like) operation as reported in~\cite{Peddinti2024}. Furthermore, the mapping from the computational domain $\Omega_0$ to the physical domain $\Omega$ must again be taken into account, as detailed in Appendix~\ref{app:stresses}.

Figure~\ref{fig:lift_oscillation} presents the temporal evolution of $C_{\text{L}}$ over 50 passage times for  $Re = 100, 150$, and $200$. The figure compares the results obtained from the TT approach on the medium grid $\mathcal{G}2$ for two different bond dimensions, i.e.,~$\chi=40$ (orange) and $\chi=60$ (magenta), and the FD method (gray). The  TT results clearly capture distinct oscillations. 
However, the agreement between $\chi=40$ and FD slightly deteriorates as $Re$ increases. Notably, the phase seems to be shifting during the simulation of the higher $Re$ number in combination with $\chi=40$. 
This is attributed to an expected increase of the required bond dimension with $Re$ at the same  accuracy, as also seen in Ref.~\cite{Kiffner2023}.

To study how the observed deterioration in Fig.~\ref{fig:lift_oscillation} affects the accuracy of the predicted Strouhal numbers, the angular frequencies of the lift forces were evaluated for a range of bond dimensions for the two coarser grids $\mathcal{G}1$ and $\mathcal{G}2$, respectively.
Results were obtained from successive (interpolated) zero crossings of \(C_{\mathrm L}\) over 50 passage times ($t \in [50, 100]$); these crossings were then used to compute the averaged shedding period.
\begin{figure}[htbp]
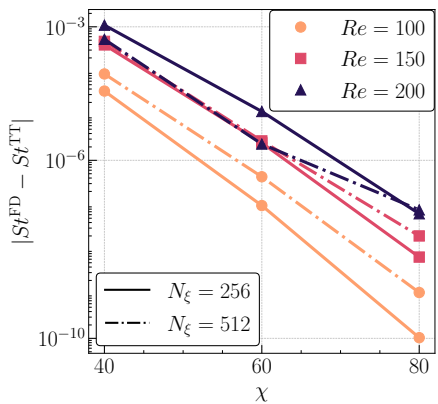

\include{Figures/TransientSec/fig_12_tikz_code}
    \caption{
    Absolute difference of the Strouhal number $St$ predicted by the TT and the classical FD method for the flow around a non-rotating cylinder on the two coarser grids $\mathcal{G}1$ (straight)  and $\mathcal{G}2$ (dot-dashed).
    }
   \label{fig:St_abs_error}
\end{figure}
\noindent
The difference between Strouhal numbers obtained from the TT simulations and the corresponding FD are presented in Fig.~\ref{fig:St_abs_error} for bond dimensions $\chi = 40,\, 60$ and $\chi = 80$.
As a point of reference, we obtain $St \approx 0.1916$ for $Re = 200$ from the FD results using $\mathcal{G}2$ which presents a $10^{-3}$ absolute error w.r.t.~the value of $St \approx 0.1977$ reported in Ref.~\cite{Ferziger2020}. 
This allows us to conclude that the prediction by TT is indeed generally very accurate, as confirmed by Fig.~\ref{fig:St_abs_error}.
Results demonstrate that as $\chi$ increases, the difference in the predicted Strouhal number decreases monotonically, consistent with the observed decrease in pointwise field differences shown in Fig.~\ref{fig:snapshots_error_distr}.
Moreover, we observe the expected trend where, for a fixed system size and bond dimension, the error in $St$ increases with growing $Re$.
Despite this, the TT-based model is particularly promising, as increases in the bond dimension $\chi$ can provide increasingly accurate Strouhal number estimates, which may be used in a Richardson extrapolation scheme to approximate the exact value.

\subsection{Rotating Cylinder}
\label{sec:rot_cylinder}
As a further demonstration of the framework’s versatility, we examine the case of a cylinder rotating with the non-dimensional rotation rate $\dot{{\theta}}= \dcyl\,\dot{\tilde{\theta}}/(2\, U_\infty) = 1$, which gives rise to the widely studied phenomenon known as the Magnus effect.
To this end, the boundary condition of the cylinder is changed to a no-slip condition with a non-vanishing prescribed tangential velocity 
\begin{equation}
    U_\Gamma=\dot{\theta} U_\infty \, .
\end{equation}
 The components of a normalized velocity $\vecc{v}_{\Gamma}$ at the discrete surface points are given by
\begin{equation}
\begin{split}
    u_{\Gamma_j} &= \dot{\theta} \cdot  \sin(2\pi j\, \delta \xi) \qquad \text{and} \\
    v_{\Gamma_j} &= -\dot{\theta}  \cdot  \cos(2\pi j\, \delta \xi) \, , 
\end{split}\label{eq:rot_bc}
\end{equation}
where $j = 0, \ldots, N_\xi - 1$.
The initial conditions are set to a uniform flow field $\vecc{v}_{\text{init.}}=(U_\infty,0)^\intercal$ with no added perturbation, as employed in Section~\ref{subsec:Steady}, since the asymmetry of the boundary conditions is sufficient to trigger the vortex shedding at $\dot{\theta}=1$~\cite{MITTAL_KUMAR_2003}.

The application of Eq.~\eqref{eq:rot_bc} realizes a counter-clockwise rotation. As a consequence of the rotation, the cylinder is subject to time-averaged circulation, which mainly results in a shift of the oscillating lift force. Figure~\ref{fig:magnus_alpha_comparison_oscillation} shows the temporal evolution of the lift-force coefficient $C_\text{L}$~\eqref{eq:liftdrag} for $\dot{\theta}=1$ (magenta) in comparison to the non-rotating case ($\dot{\theta}=0$, orange) predicted by the TT method for $Re=100$ on the coarse grid $\mathcal{G}1$.
\begin{figure}[b]
  \centering
    \include{Figures/RotatingCylinder/fig_13_tikz_code}
   \label{fig:magnus_alpha_comparison_oscillation}
\end{figure}
The maximum velocity is estimated with $\gamma=3$, i.e.,~$U_{\max}= \gamma U_\infty$, so that the resulting Courant number is below unity. Accordingly, the employed time step refers to $\delta t = \tilde \delta t U_\infty/D = 2 \times 10^{-3}$.
The TT-predicted time-averaged lift force $\overline{C}_\text{L}$ and the lift Strouhal number are outlined in Table~\ref{tab:magnus} for three different bond dimensions $\chi$ = $40$, $50$, and $60$. 
\begin{table}[t]
\centering
\caption{Comparison of $\overline{C}_\text{L}$ and lift $St$ values for the flow around a rotating cylinder at $Re=100, \ \dot{\theta}=1$. Values reported by~\citeauthor{kang_choi_lee_1999}~\cite{kang_choi_lee_1999} (bold) are compared with values obtained using the TT method with different bond dimensions $\chi$ and the classical FD solver.
}
\begin{tabular}{@{}lccccc@{}}
\toprule
Discretization            & $\chi$ & \multicolumn{2}{c}{$\overline{C}_\text{L}$} & \multicolumn{2}{c}{$St$} \\
              (interior)             &       & abs                        & \%             & abs              & \%    \\ \midrule
\textbf{241 $\times$ 241} & --      & \textbf{2.4881}          & --              & \textbf{0.1655}  & --     \\

256 $\times$ 256          & 40      & 2.4919                   & 0.15           & 0.1662          & 0.42  \\
256 $\times$ 256          & 50       & 2.4908                    & 0.11           & 0.1663           & 0.49  \\
256 $\times$ 256          & 60      & 2.4908                    & 0.11           & 0.1663           & 0.49  \\
256 $\times$ 256 & FD      & 2.4908           & 0.11             & 0.1663  & 0.49     \\
\bottomrule
\end{tabular}
\label{tab:magnus}
\end{table}
These predictions compare favorably with our FD-predicted and literature-reported values extracted from~\cite{kang_choi_lee_1999, MITTAL_KUMAR_2003} which were obtained for a similar resolution.
\begin{figure}[t!]
 \centering
    \include{Figures/RotatingCylinder/fig_14_tikz_code}
   
   \label{fig:magnus_large}
\end{figure}

It is noteworthy, that the TT approximation used for the low relative error of $0.15\%$ for $\overline{C}_\text{L}$ and  $0.42\%$ for the Strouhal prediction just uses $\chi=40 \sim 30\%$ of NVPS on the coarse grid $\mathcal{G}1$. A convergence towards the corresponding FD solver is observed as $\chi$ increases, whereas convergence toward literature-reported values is not observed, which might be attributed to differences in the simulation methods and grids.
Nonetheless, we conclude that the TT-predicted values exhibit excellent agreement with the literature while reducing the computational effort at least by a third. 

Figure~\ref{fig:magnus_large} shows the corresponding velocity magnitude fields predicted by the FD solver (a,b) and the TT solver with $\chi=40$ (c,d), and the respective pointwise error distribution (e,f) analogously to Fig.~\ref{fig:snapshots_error_distr}.
The depicted snapshots clearly describe the formation of the wake and include the initial transient state ($t \approx 4$)
and the final periodic state ($t \approx 15$).
The deflection of the flow is visible and differs from the structure in Fig.~\ref{fig:Snapshots_Karman}, which is due to the rotation of the cylinder. We observe
well-defined vortices are convected downstream in the wake region. An increase in errors is detected in time, which we attribute to the accumulation of numerical noise and truncation errors in the TT format.

\subsection{Computational Characteristics \& Performance}
\label{sec:Performance}
In this section, we examine the computational effort of the algorithm.
We confine the analysis to the assessment of the transient, non-rotating cylinder case at $Re=150$ (cf.~Section~\ref{subsec:unsteady}) and assume that changes of the physical model are not affecting the parameterization of the numerics.
\bigskip

\begin{figure}[t]
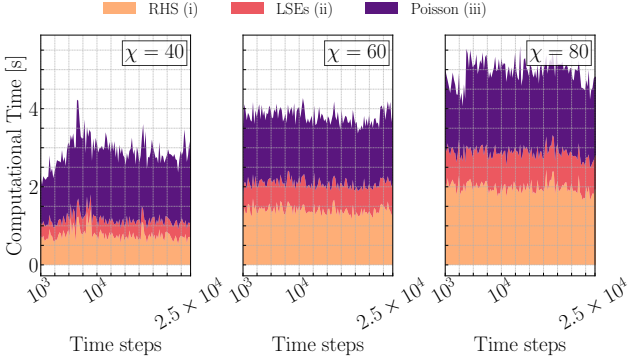

  \centering
    \include{Figures/ComputationalTime/fig_15_tikz_code}
   \caption{
   Illustration of the computing time of the three most expensive algorithmic tasks per time step for the flow around a cylinder at $Re=150$ on the coarse grid $\mathcal{G}1$ for three different bond dimensions.
   }
   \label{fig:comp_efforts}
\end{figure}

Following the fractional step approach introduced in Section~\ref{sec:algorithm}, the three most computationally expensive operations are identified. These are (i) computing the RHS of the rearranged momentum Eqs.~\eqref{eq:intermediate_vel_eq}, involving elementwise multiplications of TT-vectors to evaluate the convection term, which scales as $\mathcal{O}(\chi^{4})$, (ii) solving an LSE (cf.~Eq.~\eqref{eq:intermediate_vel_eq}) for both components of $\vecc{v}^*$, scaling as $\mathcal{O}(\chi^{3})$, and (iii) solving the Poisson Eq.~\eqref{eq:Poisson} with $\mathcal{O}(\chi^{3})$. The given scalings are taken from Table~\ref{tab:complexity}. 
The breakdown of the absolute computational effort required by these three most expensive operations is evaluated over $2.4\times 10^4$ time steps and displayed in Fig.~\ref{fig:comp_efforts}.

As evident in Fig.~\ref{fig:comp_efforts}, each computational effort (i)-(iii) grows with rising $\chi$ while the ratios between the operations remain nearly constant over time.
The results of Fig.~\ref{fig:comp_efforts} indicate that the wall-clock time per time step remains nearly constant throughout the simulation.
We attribute this behavior to the Poisson preconditioner, which enables a rapid---if not fully optimal---pressure projection even when the Poisson system is ill-conditioned and the RHS is noisy.

In contrast, an unpreconditioned DMRG solver performs less effectively with the truncation and rounding errors that accumulate over time, leading to an increase in the number of DMRG sweeps and, consequently, a higher cost per step.

With a focus on the individual operations, solving the linear system~(ii), for $\vecc{v}^*$ requires the least computational effort, significantly less than solving the Poisson problem~(iii), despite both being LSEs (cf.~Fig.~\ref{fig:comp_efforts}). This discrepancy can be attributed to the better conditioning of the left-hand side matrix \(\left(1 + \frac{\delta t}{Re} \nabla^2 \right)\) in the momentum update, resulting in faster convergence in the variational optimization. Moreover, we observe that computing the RHS~(i) via elementwise multiplications of TT-vectors has roughly the same computational effort as solving the Poisson problem~(iii).

\begin{figure}[t]
 \centering
    \includegraphics[width=\columnwidth]{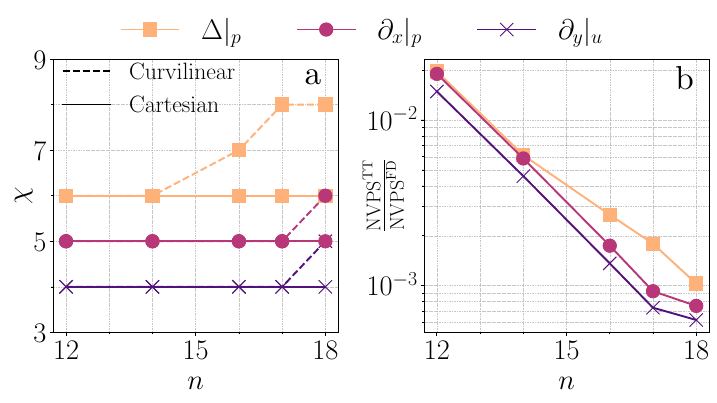}
     \caption{Compressibility analysis of the generalized TT-operators. Panel (a) depicts the required bond dimension of the curvilinear and Cartesian operators. Panel (b) shows the ratio of NVPS needed to represent the curvilinear operators in TT and sparse format.
    Please note that the lines connecting the data points are added solely for better visualization of the emerging trends.}
    \label{fig:mpo_coeffs}
\end{figure}

All these operations scale with the bond dimension $\chi$ of the fluid field representations. However, it is important to emphasize that the overall complexity also depends on the bond dimensions of the TT-matrices representing the differential operators (cf.~Table~\ref{tab:complexity}). Accordingly, Fig.~\hyperref[fig:mpo_coeffs]{\ref{fig:mpo_coeffs}\,a} illustrates the bond dimension of the curvilinear operators as TT (dashed line) w.r.t.~the corresponding operators defined on a Cartesian mesh (solid line). As it is evident in this figure, the generalized operators require a comparable bond dimension, increasing only slightly for larger grid sizes. 
Furthermore, the number of NVPS of the TT ($\mathrm{NVPS}^{\text{TT}}$) exhibits a remarkable reduction -- up to three orders of magnitude -- compared to the conventional sparse matrix representation (cf.~Fig.~\hyperref[fig:mpo_coeffs]{\ref{fig:mpo_coeffs}\,b}).
For this comparison, we consider a sparse matrix size of \(\mathrm{NVPS}^{\text{FD}}= 5N \) that represents the central point and its four neighboring points in the second-order central FD approximation of the Laplacian operator. Diagonal neighbors are not considered, due to the orthogonality of the grid.
\begin{figure}[b]
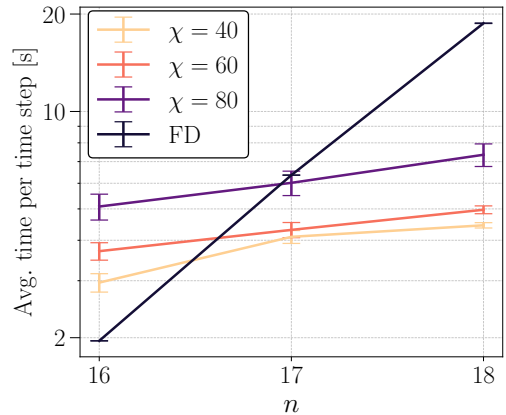

  \centering
\include{Figures/ComputationalTime/fig_17_tikz_code}
   \caption{Scaling of the average wall-clock time per iteration (sampled every $200^{\text{th}}$ iteration) for the $Re = 150$ simulations discussed in Subsection~\ref{subsec:unsteady}.}
   \label{fig:comp_times}
\end{figure}

The wall-clock time of our TT-based approach is compared to the classical FD in Fig.~\ref{fig:comp_times}. As it is depicted in this figure, both approaches reach similar computational times for the given system sizes. The slopes of the TT simulations remain constant and the growth in bond dimension $\chi$ for increasing system size is only reflected in a $y$-axis offset. On the other hand, the computational complexity of the classic FD scales exponentially with the number of tensors (TT-cores) $n$. Despite the different LSE solving strategies applied to TT and FD, such as the non-variational approach or the absence of preconditioning in the FD, the extracted conclusions from Fig.~\ref{fig:comp_times} remain valid. An improvement in the FD computation times would shift the break-even point along the $x$-axis toward higher tensor counts. Given that industrial CFD cases routinely exceed 20 million NVPS ($\approx$ 24 tensors), we anticipate that tensor-network methods like the one proposed here could deliver substantial computational gains.  

\section{Conclusions \& Outlook}
\label{sec:conclusions}
In this work, we demonstrated that the TT format can serve as an efficient ROM framework for simulating INSE in curvilinear coordinates. To the best of our knowledge, this study presents the first application of tensor network methods to the encoding and simulation of fluid fields on non-uniform, body-fitted meshes, extending their use beyond the uniform discretizations commonly considered in previous works~\cite{Kiffner2023, Peddinti2024, Kornev2023}. This development marks a significant advancement toward the realization of an industrial-grade Quantum Computational Fluid Dynamics toolbox, as body-fitted meshes are integral to many established CFD programs~\cite{Visbal_2002, Deck_2018, Ferziger2020}.

The methodology presented here ranges from outlining the mesh generation process, formulated within the TT format, to describing the curvilinear transformation of differential operators using known TT operations, and includes the efficient readout of relevant quantities of interest, such as the lift coefficient and the Strouhal number, directly in the TT format~\cite{Peddinti2024}. 
Furthermore, the approach is reinforced through the ghost point strategy for boundary treatment that eliminates the need for penalty terms in the system of equations. 
This enables the versatile application of common engineering boundaries within the TT format. Similarly, the fractional-step method avoids determining additional degrees of freedom required to enforce the pressure-velocity coupling as realized in Refs.~\cite{Gourianov2022,Hoelscher2024,Peddinti2024}. 

This allows the implementation of various flow scenarios around an immersed cylinder, ranging from Stokes flows to vortex streets with instabilities at large Reynolds numbers, and even the Magnus effect induced by cylinder rotation. In all cases, the TT-based solver demonstrates outstanding agreement
with a significant reduction up to 20-fold of the NVPS required to represent fields and up to 1000-fold for the operators.

This work lays the foundation for future investigations into how obstacle geometry, mesh generation, upwind-biased schemes~\cite{Ye2022,Danis2025,Bengoechea2024}, parallelization~\cite{Secular2020}, and grid properties (e.g.,~orthogonality) influence the achievable compression of operator representations and further increase the TT's efficiency. In particular, certain mesh types--such as elliptic grids--may naturally yield low-rank, geometry-aware operators, offering potential for shape optimization and other industrial applications. A natural next step is to investigate the laminar and turbulent flow around an immersed ellipsoid, which breaks the rotational symmetry seen in the cylinder, as well as the flow around NACA airfoils~\cite{Prandtl1923}, providing a compelling test case. Particular focus will be placed on high-Reynolds number regimes, where the curvilinear approach is expected to be advantageous compared to IBM.

\begin{acknowledgments}
This publication and the current work have received funding from the European Union's Horizon Europe research and innovation program (HORIZON-CL4-2021-DIGITAL-EMERGING-02-10) under grant agreement No. 101080085 QCFD.
PS acknowledges financial support by the DLR-Quantum-Fellowship Program.
DJ and TH are partly funded by the Cluster of Excellence `Advanced Imaging of Matter' of the Deutsche Forschungsgemeinschaft (DFG)|EXC 2056- project ID390715994.
DJ acknowledges the support by DFG project ``Quantencomputing mit neutralen Atomen'' (JA 1793/1-1, Japan-JST-DFG-ASPIRE 2024) and  the Hamburg Quantum Computing Initiative (HQIC) project EFRE. The EFRE project is co-financed by ERDF of the European Union and by ``Fonds of the Hamburg Ministry of Science, Research, Equalities and Districts (BWFGB)''. The numerical data is acquired using the computational cluster Hummel-2 of the University of Hamburg, which is funded by the Deutsche Forschungsgemeinschaft (DFG, German Research Foundation) – 498394658. This work was inspired in part by the engaging discussions and expert input from Airbus, BMW Group, and AWS during the \textit{Quantum Mobility Quest} challenge. 
The authors thank Philipp~Thoma for fruitful discussions.
\end{acknowledgments}

\section*{Author Declarations}
\addcontentsline{toc}{section}{Author Declarations}

\subsection*{Declaration of Interest}
\addcontentsline{toc}{subsection}{Declaration of Interest}
The authors have no interest to disclose. 

\subsection*{Author Contributions}
\addcontentsline{toc}{subsection}{Author Contributions}
\textbf{Nis-Luca~van~H\"ulst:} Conceptualization, Data curation, Formal analysis, Investigation, Methodology, Software, Validation, Visualization, Writing – original draft, Writing – review and editing; \textbf{Pia~Siegl:} Conceptualization, Data curation, Formal analysis, Investigation, Methodology, Software, Visualization, Writing – original draft, Writing – review and editing; \textbf{Paul~Over:} Conceptualization, Data curation, Formal analysis, Investigation, Methodology, Visualization, Writing – original draft, Writing – review and editing; \textbf{Sergio~Bengoechea:} Conceptualization, Data curation, Formal analysis, Investigation, Methodology, Visualization, Writing – original draft, Writing – review and editing; \textbf{Tomohiro~Hashizume:} Conceptualization, Investigation, Methodology, Supervision, Visualization, Writing – original draft, Writing – review and editing; \textbf{Mario~Guillaume~Cecile:} Conceptualization, Investigation, Methodology, Supervision, Visualization, Writing – original draft, Writing – review and editing; \textbf{Thomas~Rung:} Project administration, Funding acquisition, Supervision, Conceptualization, Methodology, Resources, Writing - review and editing; \textbf{Dieter~Jaksch:} Project administration, Funding acquisition, Supervision, Conceptualization, Methodology,  Resources, Writing - review and editing. 

\section*{Data availability}
A software example for simulating the 2D heat equation on the employed cylindrical grid is available via
Ref.~\cite{hulstasoftware2025}.
The data that support the findings of this study are available upon reasonable request from the authors and at Ref.~\href{https://doi.org/10.25592/uhhfdm.17687}{https://doi.org/10.25592/uhhfdm.17687}.

\bibliography{apssamp}

\clearpage
\newpage

\onecolumngrid
\appendix

\section{Discretization and Coordinate Transformation}
\label{App:DiscretizationCurvis}
\begin{figure}[b]
    \centering
    \begin{minipage}[]{0.42\textwidth}
        \centering
        \def\svgwidth{\linewidth}
        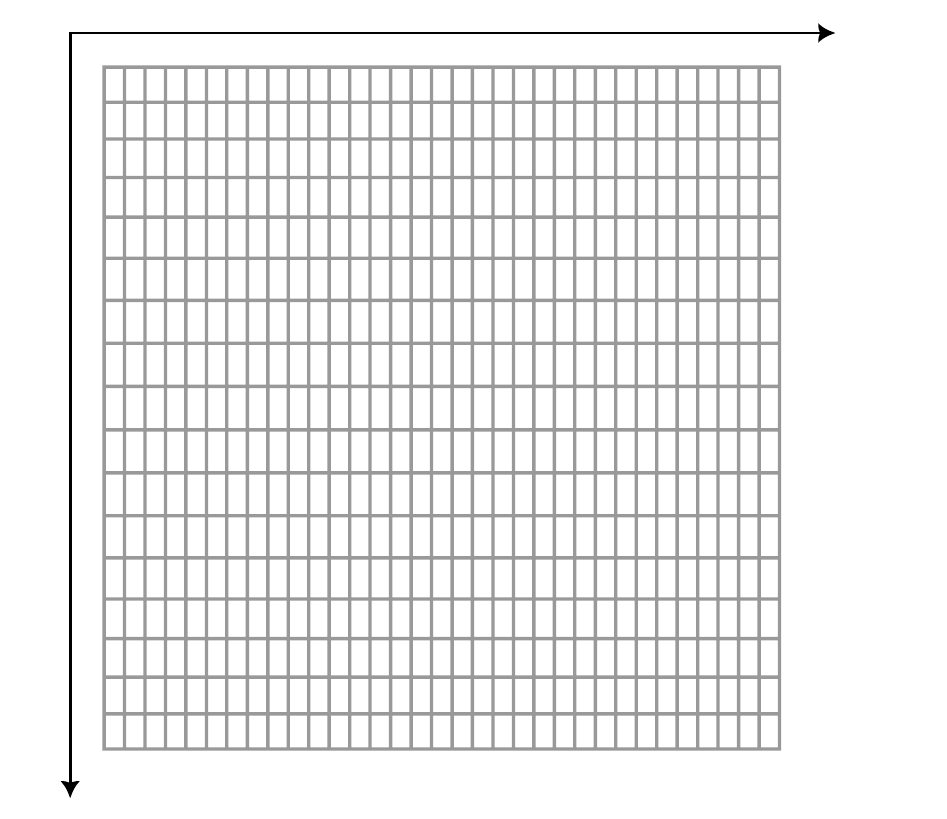
    \end{minipage}
    \hfill
    \begin{minipage}[]{0.48\textwidth}
        \centering
        \def\svgwidth{\linewidth}
        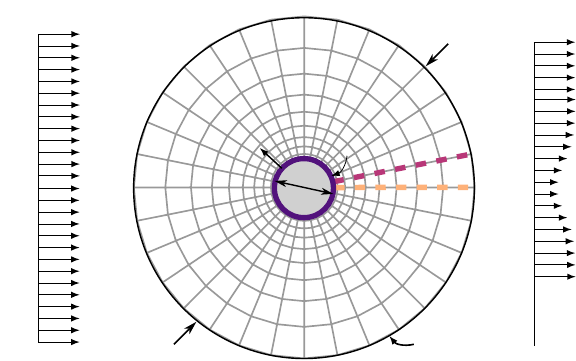
    \end{minipage}

    \caption{Computational domain \( \Omega_0 \) with uniform discretization (left)
and physical domain \( \Omega \) around a cylinder discretized using an O-type grid (right). 
In both, light-gray points indicate grid nodes, and colored lines highlight corresponding boundaries.}
    \label{fig:mapping}
\end{figure}

In Fig.~\ref{fig:mapping} the two domains of interest are illustrated, i.e.,~the computational (a) and the physical (b) domain. The former corresponds to a uniformly discretized unit square \( \Omega_0 = [0, 1] \times [0, 1) \), while the latter is discretized using an O-type grid. The coloring highlights the corresponding boundaries of the two domains; for instance, the outer boundary \( \partial \Omega \) of the physical domain (shown in black) maps to the top boundary of the unit square (also black).

The computational domain \( \Omega_0 \) is discretized using \( (2^{n_\eta} + 2) \times 2^{n_\xi} \) grid points in the \( \eta(i) \)- and \( \xi(j) \)-directions, respectively, given by
\begin{equation}
   \label{eq:discretization_unit_square}
    \begin{rcases2}
  \eta_{ij}&=(i+1) \delta \eta \  \\
  \xi_{ij}&=j \delta \xi \ 
    \end{rcases2} \, 
\text{with } 
    \begin{array}{l}
        i = -1, \ldots, 2^{n_\eta} \ , \\
        j = 0, \ldots, 2^{n_\xi} - 1 \ . 
    \end{array}
\end{equation}
Here, \( i = -1 \) and \( i = 2^{n_\eta} \) correspond to ghost points representing the top (black) and bottom (magenta) boundaries, respectively. These grid points are not part of the solution domain but are incorporated through the ghost-point approach.
The associated grid spacings are \( \delta \eta = 1 / (2^{n_\eta}+ 1) \) and \( \delta \xi = 1 / 2^{n_\xi} \). The difference in spacing reflects the boundary conditions: periodicity in the \( \xi \)-direction versus non-periodicity in \( \eta \).

The Cartesian spatial derivatives required by the momentum~\eqref{eq:Navier-Stokes-Eq-B} and continuity \eqref{eq:Conti-Eq-B} equations (INSE) are expressed in the reference coordinate system $(\xi, \eta)$ by applying the chain rule, e.g.,~$
\partial_x = \partial_x\xi \, \partial_\xi + \partial_x\eta \,   \partial_\eta$, and using a coordinate transformation, i.e.,~$\partial_x \xi=\partial_\eta z/J$, $\partial_x\eta=-\partial_\xi z/J$, $\partial_y \xi=-\partial_\eta x/J$ and $\partial_y \eta=\partial_\xi x/J$, viz.
\begin{align}
\label{eq:curviliner-deriavtive}
    (\partial_x, \partial_z) = \frac{1}{J}
    \begin{pmatrix}
        \partial_\eta z & -\partial_\xi z \\
        -\partial_\eta x & \partial_\xi x
    \end{pmatrix}
    (\partial_\xi,  \partial_\eta)^\intercal\, ,
\end{align}
for first spatial derivatives and 
\begin{align}
    \Delta =  \frac{1}{J}(\partial_\xi, \partial_\eta)
     \frac{1}{J} \begin{pmatrix}
         g_{22} &  -g_{12} \\
         -g_{21}  & g_{11}
    \end{pmatrix}  (\partial_\xi,  \partial_\eta)^\intercal\, ,
\end{align}
for the Laplace operator. Here, $J$ denotes the Jacobian determinant and $g$ the metric tensor. They are given by the following expressions
\begin{equation}
    \begin{split}
        J = (\partial_\xi x)(\partial_\eta z) - (\partial_\eta x)(\partial_\xi z) 
    = \det \begin{pmatrix}
    \partial_\xi x & \partial_\eta x \\
    \partial_\xi z & \partial_\eta z
    \end{pmatrix} \ &,  \\
    g_{11} = (\partial_\xi x)^2 + (\partial_\xi z)^2 \ ,\quad   
    g_{22} = (\partial_\eta x)^2 + (\partial_\eta z)^2 \ , \quad 
    g_{12} = (\partial_\xi x)(\partial_\eta x) + (\partial_\xi z)(\partial_\eta z)\ & .
    \end{split}
\end{equation}
Writing out the Laplacian operator one obtains a sum of seven operator terms
\begin{equation}
    \Delta = \frac{1}{J} \Bigg\{
\partial_\xi\left( \frac{g_{22}}{J} \right) \partial_\xi 
+ \frac{g_{22}}{J} \, \partial_{\xi\xi} 
- \partial_\xi\left( \frac{g_{12}}{J} \right) \partial_\eta 
- \partial_\eta\left( \frac{g_{12}}{J} \right) \partial_\xi 
- 2 \frac{g_{12}}{J} \, \partial_{\xi\eta} 
+ \partial_\eta\left( \frac{g_{11}}{J} \right) \partial_\eta 
+ \frac{g_{11}}{J} \, \partial_{\eta\eta}
\Bigg\} \, .
\end{equation}
The derivatives w.r.t. $\xi$ and $\eta$ are approximated with second-order central difference methods. The first  derivatives of a generic scalar field \( \phi \) read 
\begin{equation}
    \begin{split}
       \partial_\xi \phi\Big|_{ij}  = \frac{\phi_{i,j+1}-\phi_{i,j-1}}{2\delta \xi} + \mathcal{O}(\delta \xi^2)\,, \quad
        \partial_\eta \phi\Big|_{ij}  = \frac{\phi_{i+1,j}-\phi_{i-1,j}}{2\delta \eta} + \mathcal{O}(\delta \eta^2)\,.
    \end{split}
\end{equation}
Similarly, the corresponding second derivatives read
\begin{equation}
\begin{split}
     \partial_{\xi\xi} \phi \Big|_{ij} = \frac{\phi_{i,j+1}-2\phi_{i,j}+\phi_{i,j-1}}{\delta \xi^2}  + \mathcal{O}(\delta \xi^2)\ , \quad
      \partial_{\eta\eta}\phi \Big|_{ij} = \frac{\phi_{i+1,j}-2\phi_{i,j}+\phi_{i-1,j}}{\delta \eta^2}  + \mathcal{O}(\delta \eta^2)\ .
\end{split}
\end{equation}
and finally mixed derivatives are approximated by 
\begin{equation}
\left.\partial_{\xi\eta} \phi \right|_{i,j}
= \frac{ \phi_{i+1,j+1} - \phi_{i+1,j-1} - \phi_{i-1,j+1} + \phi_{i-1,j-1} }{4 \delta \xi \delta \eta} + \mathcal{O}(\delta \xi^2 + \delta \eta^2)\,.
\end{equation}

\section{Sorting Vector to Matrix and Reverse}
\label{App:MatrixVectorSorting}

The column-major ordered vectorization of matrix \( A \in \mathbb{R}^{3 \times 3} \), denoted \( \operatorname{vec}(A) \), is
\begin{equation} 
    A = 
\begin{bmatrix}
a_{11} & a_{12} & a_{13} \\
a_{21} & a_{22} & a_{23} \\
a_{31} & a_{32} & a_{33}
\end{bmatrix} \quad \rightarrow  \quad \operatorname{vec}(A)= [a_{11}
a_{21} \ 
a_{31} \ 
a_{12} \ 
a_{22} \ 
a_{32}\ 
a_{13}\ 
a_{23}\ 
a_{33} ]^\intercal \ .
\end{equation}

\section{Read-out of Column or Row }
\label{App:Readout}
Assuming the two-dimensional scalar field \( f_{ij} := f(\xi_{ij}, \eta_{ij}) \in \mathbb{R}^{2^{n_\eta} \times 2^{n_\xi}} \) is represented as a TT, we write
\begin{equation}
    f_{ij} \simeq 
    \tensor{A}[1]^{i_1}
    \tensor{A}[2]^{i_2}
    \cdots
    \tensor{A}[n_\eta]^{i_{n_\eta}}
    \tensor{A}[n_\eta+1]^{j_1}
    \cdots
    \tensor{A}[n]^{j_{n_\xi}} \ ,
\end{equation}
where total number of TT-cores is \( n = n_\eta + n_\xi \). Each core \( \tensor{A}[\ell] \in \mathbb{R}^{\alpha_{\ell-1} \times 2 \times \alpha_\ell} \) is equipped with a corresponding binary digit \( i_k, j_k \in \{0,1\} \) of the row and column indices, respectively.

To extract the \( I \)-th row, i.e.,~the slice \( f_{Ij} \), we fix the first \( n_\eta \) physical indices using the binary representation \( (I_1, \dots, I_{n_\eta}) \in \{0,1\}^{n_\eta} \) of \( I \), resulting in the contracted form
\begin{equation}
    f_{Ij} \simeq 
    \tensor{A}[1]^{i_1 = I_1}
    \tensor{A}[2]^{i_2 = I_2}
    \cdots
    \tensor{A}[n_\eta]^{i_{n_\eta} = I_{n_\eta}}
    \tensor{A}[n_\eta+1]^{j_1}
    \cdots
    \tensor{A}[n]^{j_{n_\xi}} \ .
\end{equation}
This procedure effectively contracts the first \( n_\eta \) tensor cores to form a matrix \( \mat{M}_I \in \mathbb{R}^{1 \times \alpha_{n_\eta}} \), which is then left-multiplied into the remaining TT-cores
\begin{equation}
    f_{Ij} \simeq \mat{M}_I \cdot 
    \left( 
    \tensor{A}[n_\eta+1]^{j_1}
    \cdots
    \tensor{A}[n]^{j_{n_\xi}} 
    \right) \ .
\end{equation}
This yields a TT representation of the row \( f_{Ij} \in \mathbb{R}^{2^{n_\xi}} \). Similarly, to extract the \( J \)-th column, i.e.,~the slice \( f_{iJ} \), we fix the last \( n_\xi \) physical indices using the binary representation \( (J_1, \dots, J_{n_\xi}) \in \{0,1\}^{n_\xi} \) of \( J \), resulting in the contracted form
\begin{equation}
    f_{iJ} \simeq 
    \tensor{A}[1]^{i_1}
    \tensor{A}[2]^{i_2}
    \cdots
    \tensor{A}[n_\eta]^{i_{n_\eta}}
    \tensor{A}[n_\eta+1]^{j_1=J_1}
    \cdots
    \tensor{A}[n]^{j_{n_\xi}=J_{n_\xi}} \ .
\end{equation}
This procedure effectively contracts the last \( n_\xi \) tensor cores to form a matrix \( \mat{M}_J \in \mathbb{R}^{ \alpha_{n_\eta} \times 1} \), which is then right-multiplied into the remaining TT-cores:
\begin{equation}
    f_{iJ} \simeq  
    \left( 
    \tensor{A}[1]^{i_1}
    \tensor{A}[2]^{i_2}
    \cdots
    \tensor{A}[n_\eta]^{i_{n_\eta}} 
    \right) \cdot \mat{M}_J \, .
\end{equation}
This yields a TT representation of the column \( f_{iJ} \in \mathbb{R}^{2^{n_\eta}} \).

\section{Vector to Tensor Train Decomposition}
\label{App:TensorTrain}
The decomposition of a vector into a TT is exemplified here analytically for a vector of size $2^3=8$ ($n=3$). 
As a preliminary requirement, mention the rank core product, denoted by $\Join$, which acts as an array multiplication given vector entities as 
\begin{equation}
 \begin{bmatrix}
        \begin{pmatrix}
            a\\
            b
        \end{pmatrix} \begin{pmatrix}
            c\\
            d
        \end{pmatrix}
    \end{bmatrix}
    \Join
    \begin{bmatrix}
        \begin{pmatrix}
            \alpha \\
            \beta 
        \end{pmatrix}\\
        \begin{pmatrix}
            \gamma\\
            \delta
        \end{pmatrix}
    \end{bmatrix} = \begin{pmatrix}
        a\\
        b
    \end{pmatrix} \otimes \begin{pmatrix}
    \alpha \\
    \beta 
    \end{pmatrix} + \begin{pmatrix}
        c \\
        d
    \end{pmatrix} \otimes 
    \begin{pmatrix}
        \gamma \\
        \delta 
    \end{pmatrix}  = \begin{pmatrix}
        a\alpha + c\gamma \\
        a\beta + c\delta\\
        b\alpha + d\gamma\\
        b\beta + d\delta
    \end{pmatrix}\ ,
\end{equation}

\begin{equation}
    \begin{bmatrix}
        \begin{pmatrix}
            a\\
            b \\
            c \\
            d
        \end{pmatrix} \begin{pmatrix}
            e\\
            f \\
            g\\
            h
        \end{pmatrix}
    \end{bmatrix}
    \Join
    \begin{bmatrix}
        \begin{pmatrix}
            \alpha \\
            \beta 
        \end{pmatrix}\\
        \begin{pmatrix}
            \gamma\\
            \delta
        \end{pmatrix}
    \end{bmatrix} = \begin{pmatrix}
        a\\
        b\\
        c\\
        d
    \end{pmatrix} \otimes \begin{pmatrix}
    \alpha \\
    \beta 
    \end{pmatrix} + \begin{pmatrix}
        e\\
        f\\
        g \\
        h
    \end{pmatrix} \otimes 
    \begin{pmatrix}
        \gamma \\
        \delta 
    \end{pmatrix}  = \begin{pmatrix}
        a\alpha + e\gamma \\
        a\beta + e\delta\\
        b\alpha + f\gamma\\
        b\beta + f\delta \\
        c\alpha + g\gamma \\
        c\beta + g\delta\\
        d\alpha + h\gamma\\
        d\beta + h\delta
    \end{pmatrix} \ .
\end{equation}

If one now defines a vector $\vecc{t}=\left(1,2,3,4,5,6,7,8\right)^\intercal$, its tensor decomposition is
\begin{align}
\begin{bmatrix}
        \begin{pmatrix}
            1\\
            1
        \end{pmatrix} & \begin{pmatrix}
            0\\
            1
        \end{pmatrix}
    \end{bmatrix}
    \Join
    \begin{bmatrix}
        \begin{pmatrix}
            1\\
            1
        \end{pmatrix} & \begin{pmatrix}
            0\\
            1
        \end{pmatrix}\\
        \begin{pmatrix}
            0\\
            0
        \end{pmatrix} & \begin{pmatrix}
            2\\
            2
        \end{pmatrix}
    \end{bmatrix} \Join
    \begin{bmatrix}
        \begin{pmatrix}
            1\\
            2
        \end{pmatrix}\\
        \begin{pmatrix}
            2\\
            2
        \end{pmatrix}
    \end{bmatrix} =     \begin{pmatrix}
        1\\
        2\\
        3\\
        4\\
        5\\
        6\\
        7\\
        8\\
    \end{pmatrix}.
\end{align}
Each of the square brackets represents one tensor core $\tensor{A}[k]$, where $k\in\{0,1,2\}$. Furthermore, the bond dimension and the physical dimension are $2$ in all cases.

\section{Grid points in Application}
\label{app:gridpoints}
 The classically generated grid consists of $ N = N_\eta \times N_\xi := 2^{n_\eta} \times 2^{n_\xi}$ interior support points, with indices $i = 0, \ldots, N_\eta - 1$ and $ j = 0, \ldots, N_\xi - 1 $, provided as a tensor (outer) product of 1D functions as
\begin{equation}
    \label{eq:mesh}
    x_{ij} = \frac{\dcyl}{2} \left( \frac{D_{\partial \Omega}}{ \dcyl} \right)^{\frac{i+1}{N_\eta + 1}} \cos(2 \pi j \delta \xi) \ , \quad 
    z_{ij} = \frac{\dcyl}{2} \left( \frac{D_{\partial \Omega}}{ \dcyl} \right)^{\frac{i+1}{N_\eta + 1}} \sin(2 \pi j \delta \xi) \ .
\end{equation}

\section{Force Computation}
\label{app:stresses}
The Strouhal number can be computed using different methods, depending on the available data and flow characteristics. In this study, we determine the Strouhal number from the oscillations in the  dimensionless lift force $F_\text{L}$,  following the approach of~\cite{Ferziger2020}. The lift force corresponds to the $z$-projection of the total force as
\begin{equation} 
     F_\text{L} = \vecc{e}_z^\intercal  \int_\Gamma \matc{\tau}(\vecc{v},p) \vecc{n} \,  ds\ \label{eq:lift_force}
    \approx   2\pi \delta \xi  \frac{1}{2}\sum_{j=0}^{N_\xi-1} (\tau_{21}, \tau_{22}) 
    \begin{pmatrix}
        \cos{2\pi j \delta \xi} \\
        \sin{-2\pi j \delta \xi}
    \end{pmatrix}\ .
\end{equation}
Here, the stress tensor components need to be determined on the cylinder surface via
\begin{equation}
     \tau_{21} = \frac{2}{Re}\Bigg(\frac{1}{2}\left( \frac{\partial v}{\partial x}\Big|_\Gamma + \frac{\partial u}{\partial z}\Big|_\Gamma \right) \Bigg) \ , \quad
    \tau_{22} = - \left( 
    p\Big|_{\Gamma} + \frac{2}{3Re}(\nabla \vecc{v})\Big|_{\Gamma}  \right) + \frac{2}{Re}\frac{\partial v}{\partial z}\Big|_{\Gamma}\ ,
\end{equation}
where derivatives are computed in the computational domain $\Omega_0$ and then transformed. A similar explicit formula can be derived for the imensionless drag force $F_D$.

\section{Algorithm Modifications for Transient}
\label{app:AlgorithmModificationsforTransient}
First, we implement an initial modulation onto the velocity field to allow the vortex shedding to emerge. To do so, we employed the following modulation
\begin{equation}
    u_{ij}(t=0) = A_0 \frac{e^{(i \delta \eta)^6} - 1}{e - 1} \sin{\left(2\pi j \delta\xi \right)}\ ,\label{eq:init_cond}
\end{equation}
with perturbation strength $A_0=0.05$. This modulation introduces an asymmetry w.r.t. the branch cut, while it vanishes radially towards the far-field boundary $\partial \Omega$. This was motivated to avoid strong perturbations on the comparably large cells in the far field, potentially making the system stiffer.
Second, we employ a preconditioner for solving the Poisson equation within the TT approach to enable faster convergence in terms of the number of sweeps in the variational scheme~\cite{Schollwoeck2011}, and to avoid dealing with highly ill-conditioned LSE in the local variational updates. This allows us to make use of the \textit{fast matvec} routine outlined in~\cite{Dolgov2012}, i.e.,~emulating the local matrix-vector multiplication by successive tensor contractions scaling with $O(\chi^3)$ instead of assembling the local matrix and naively applying it to a vector with scaling of $O(\chi^4)$.

\section{Verification of the Finite Difference Reference}
\label{app:openfoam}
\begin{figure}[htbp]
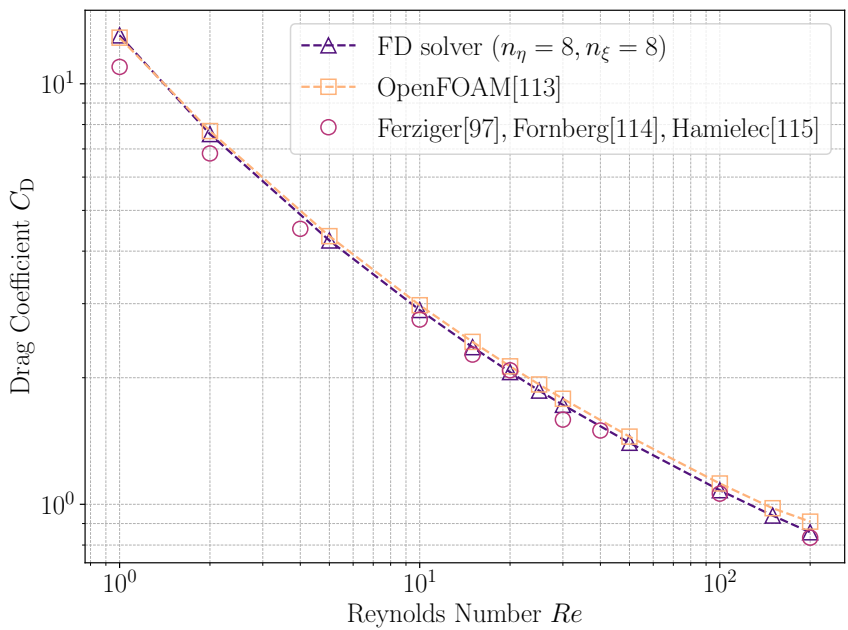

    \centering
    \include{Figures/validation_appendix}
    \caption{Drag coefficient as a function of the Reynolds number $Re$ for the flow around a cylinder from Stokes to subcritical regimes. The benchmark study compares results from literature~\cite{Ferziger2020,Hamielec1969,FORNBERG1985}, OpenFOAM~\cite{engys} and the developed FD solver.}
    \label{fig:benchmark_cyl}
\end{figure}

To validate the feasibility of the presented results, the in-house FD code is benchmarked against reference values from OpenFOAM~\cite{engys} and the literature~\cite{Ferziger2020,FORNBERG1985,Hamielec1969} for the case of flow around a cylinder. For a quantitative comparison, the drag coefficient $C_\text{D}$ as a function of the Reynolds number over the range $Re \in [1,200]$ is given in Fig.~\ref{fig:benchmark_cyl}. 

The verification uses $\bigcirc$ marks for literature references, $\square$ marks for the numerical results from OpenFOAM~\cite{engys}, and $\triangle$ for results of the FD solver under validation. Overall, the results confirm good agreement, though minor discrepancies are apparent, especially at lower $Re$. These deviations can be attributed mainly to two factors: 1) discretization differences between the numerical models (finite approximation, grid resolution) and 2) differences in the computational setup (boundary layers, pressure correction approach, boundary conditions).

1) The application of different meshes between the models in Fig.~\ref{fig:benchmark_cyl} is a source of discrepancy in the results. 
Furthermore, the references either approximate the vorticity-streamfunction formulation in \cite{Hamielec1969,FORNBERG1985}, 
or apply finite volume discretizations \cite{engys,Ferziger2020}, which inherently conserve fluxes, whereas the in-house FD solver uses central stencils introducing dispersion errors.

2) Another source of error lies in the computational model itself. As conventional CFD solvers can handle more complex flow phenomena, different types of boundaries and pressure correction approaches are implemented. In contrast, the developed FD code relies on Dirichlet-Neumann pressure boundaries to guarantee numerical stability. The differences in the boundary conditions are important as the wall treatment directly influences the computation of forces and reflects on the drag coefficient $C_\text{D}$. As a consequence, the corresponding literature symbols do not correctly align with the numerical results that include inertial contributions. 

In conclusion, the FD solver delivers good accuracies and shows a constant offset to the given references, indicating a well-functioning of the model and motivating a reliable basis for benchmarking the TT solver under discussion. 
\end{document}